\documentclass[12pt]{article}
\usepackage{geometry}
\usepackage{amsmath, amssymb, graphicx,fullpage,color,mathtools,amsthm,xcolor}
\usepackage{caption, subcaption}
\usepackage{algorithm, algorithmic}
\usepackage{multirow, booktabs}

\graphicspath{{./figure/}}

\usepackage[normalem]{ulem}

\usepackage{hyperref,color}

\definecolor{webgreen}{rgb}{0,.35,0}
\definecolor{webbrown}{rgb}{.6,0,0}
\definecolor{RoyalBlue}{rgb}{0,0,0.9}
\definecolor{purp}{rgb}{0.6,0.05,0.8}
\definecolor{ora}{rgb}{0.7,0.35,0.02}

\hypersetup{
   colorlinks=true, linktocpage=true, 
   urlcolor=webbrown, linkcolor=RoyalBlue, citecolor=webgreen,
   pdfauthor={Shunyu Yao, Gary P. T. Choi},
   pdfsubject={Conformal Tubular Parameterization and Toroidal Bending of Tube-Like Surfaces}
}

\begin{document}
\author{Shunyu Yao\href{https://orcid.org/0009-0007-5074-4533}{\protect\includegraphics[scale=.050]{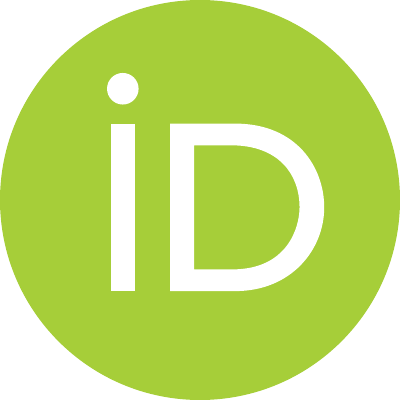}}$^{1}$,  Gary P. T. Choi\href{https://orcid.org/0000-0001-5407-9111}{\protect\includegraphics[scale=.050]{ORCID.pdf}}$^{1,\ast}$\\
\\
\footnotesize{$^{1}$Department of Mathematics, The Chinese University of Hong Kong, Hong Kong}\\
\footnotesize{$^\ast$To whom correspondence should be addressed; E-mail: ptchoi@cuhk.edu.hk}
}
\title{Conformal Tubular Parameterization and Toroidal Bending of Tube-Like Surfaces}

\date{}
\maketitle

\begin{abstract}
Tube-like surfaces are widely encountered in geometry processing, engineering structures, and medical anatomy, yet their intrinsic longitudinal and circumferential information is not well preserved by conventional planar annular or rectangular parameterization domains. In this work, we propose a new conformal parameterization framework for open tube-like surfaces with two boundary components. The proposed method first constructs a fixed-boundary tubular parameterization by cutting the input mesh, computing a disk-to-parallelogram quasi-conformal map, and lifting the result to a cylindrical surface in $\mathbb{R}^3$. To reduce residual distortion introduced near the cut seam, we further introduce a localized quasi-conformal correction scheme formulated on an annular domain, which improves conformality while leaving regions away from the seam unchanged. To handle noisy or irregular input boundaries, we also develop a free-boundary variant based on boundary extension and cycle-Laplacian smoothing, allowing the prescribed boundary constraints to be imposed on artificial outer rings rather than directly on the original surface. Finally, as an optional post-processing step, we derive two conformal toroidal bending maps that embed the tubular parameterization into toroidal geometries without altering its conformal structure. Experiments on synthetic tube meshes and real vascular surfaces demonstrate that the proposed framework produces low-distortion parameterizations, effectively mitigates seam-induced artifacts, improves robustness for boundary-noisy inputs, and provides flexible tubular and toroidal target domains for downstream surface processing tasks.
\end{abstract}

%%%%%%%%%%%%%%%%%%%%%%%%
\section{Introduction} \label{sect:introduction}

Surface parameterization maps a surface to a canonical domain and is a fundamental tool in computer graphics, geometry processing, medical visualization, and shape analysis. High-quality parameterizations facilitate tasks such as texture mapping, remeshing, shape comparison, shape morphing, and numerical computation on surfaces~\cite{zockler2000Fast,Floater2005ParaSurvey,Sheffer2006MeshPara}. Over the past several decades, a wide variety of surface parameterization methods have been developed, including barycentric and convex-combination mappings rooted in Tutte's embedding theorem~\cite{Tutte1963DrawGraph,Floater1997SmoothApprox}, stretch-minimizing and near-isometric formulations such as MIPS (Most-Isometric ParameterizationS)~\cite{Hormann2000MIPS}, intrinsic and free-boundary methods~\cite{Desbrun2002Intrinsic}, least-squares conformal formulations such as LSCM (Least Squares Conformal Map)~\cite{Levy2002LSCM}, angle-based methods such as ABF++ (Fast and Robust Angle Based Flattening)~\cite{Sheffer2005ABF++}, local/global optimization frameworks~\cite{Liu2008Local/Global}, and injective parameterizations with explicit validity guarantees~\cite{Weber2014Locally,rabinovich2017scalable}. 

Among the many available formulations, conformal parameterization is particularly attractive because it preserves angles and therefore retains local geometric structure. For instance, conformal parameterization has been widely used in texture mapping and automatic texture atlas generation~\cite{Levy2002LSCM,Haker2000Texture}, in landmark-constrained surface registration and shape analysis~\cite{Lui2010Registration}, and in medical visualization tasks such as virtual colonoscopy and colon flattening for polyp detection~\cite{Haker2000ColonCT,Hong2006Colon,zeng2010Supine,zeng2014Colon}. Beyond these application domains, several geometric-flow-based methods provide effective computational tools for constructing conformal maps. Holomorphic differentials provide a fundamental analytic tool for studying and constructing conformal maps, especially on surfaces with nontrivial topology, where they encode the underlying conformal structure and yield canonical coordinates through integration~\cite{gu2002computing,zeng20083d,kropf2014conformal}. Discrete Ricci flow computes conformal metrics through curvature prescription and has become an important framework for surface parameterization and shape analysis~\cite{jin2008discrete,zeng2013ricci}. Related evolution-based approaches, such as conformalized mean curvature flow, further improve robustness by modifying mean-curvature flow so that the evolving surface remains conformally equivalent to the input~\cite{crane2013robust,Kazhdan2012MeanCurvature}. These developments highlight the practical value of conformal maps in situations where preserving local shape and controlling angular distortion are of primary importance. For instance, in medical shape analysis, notable examples include brain surface mapping and cortical surface registration via conformal maps~\cite{Gu2004Brain,Wang2007Brain}.

Note that many recent parameterization methods have focused on simply connected open surfaces~\cite{Sawhney2017BoundaryFirst,Choi2020Parallelizable,liao2022constructive,choi2025hemispheroidal}, genus-0 surfaces~\cite{nadeem2016spherical,hu2017advanced,wang2018novel,liao2024convergence,Choi2024FastEllipsoidal,sutti2024riemannian,Lyu2024SDE}, or genus-1 surfaces~\cite{yueh2020new,Yao2026TDEM}. However, in practice, many surfaces exhibit tube topology. Representative examples include medical structures such as blood vessels, bronchial trees, intestines, and other tubular anatomical organs, as well as engineering structures such as pipelines, ducts, tunnels, and internal cooling channels. These examples suggest that tube-like surfaces are not exceptional cases but rather a common class of geometries in real-world applications. As a result, it is essential to design algorithms that can accurately represent, process, and analyze surfaces with tube topology.

A possible approach is to cut the mesh open and map it onto a planar domain with periodic boundary conditions, with notable examples including the conformal flattening methods by Haker et al.~\cite{Haker2000ColonCT}, Hong et al.~\cite{Hong2006Colon} and Marino et al.~\cite{marino2011context}, and quasi-conformal flattening methods by Zeng et al.~\cite{zeng2010Supine,zeng2014Colon} and Saucan et al.~\cite{saucan2008local}. In particular, Saucan et al.~\cite{saucan2008local} studied local and global aspects of quasi-conformal mappings for medical imaging, including mappings from annuli to rectangles.

However, the artificial cut destroys the intrinsic cyclic structure of the surface, and maintaining periodic consistency across the cut during subsequent processing requires additional constraints and numerical effort. Another natural approach is to map these surfaces onto an annulus~\cite{Choi2021multiconnect}. However, while the annulus is topologically compatible with a surface with two boundary components, flattening an elongated tubular surface onto a planar annular domain may introduce undesirable geometric distortion, particularly when the surface has a large aspect ratio or a highly nonuniform radius.

The above limitations motivate target domains that are not only topologically compatible but also geometrically aligned with the intrinsic structure of tube-like surfaces. A cylindrical or tubular parameter domain explicitly represents one longitudinal direction and one periodic circumferential direction, and therefore provides a natural coordinate system for tubular geometries. While cylinder-based parameterizations have been considered in prior work~\cite{huysmans2005Parameterization}, several issues remain important for a conformal tubular mapping framework, including the computational efficiency and mapping accuracy, as well as the handling of unreliable or irregular input boundaries.

A further possibility is to use a toroidally bent tube domain as an alternative target domain. Instead of representing the parameter space as a straight cylinder, one smoothly bends the tube into a toroidal shape, preserving the periodic circumferential structure while allowing greater geometric flexibility along the longitudinal direction. This is particularly useful for tube-like surfaces whose centerlines are naturally curved, nearly closed, or ring-like, since the target domain can better reflect their global geometry without changing the underlying tubular topology. Since such a bending can be realized by a conformal map, it does not change the intrinsic conformal structure of the tubular parameterization, but only the extrinsic geometry of the target domain.

\begin{figure}[t]
    \centering
       \includegraphics[width=\linewidth]{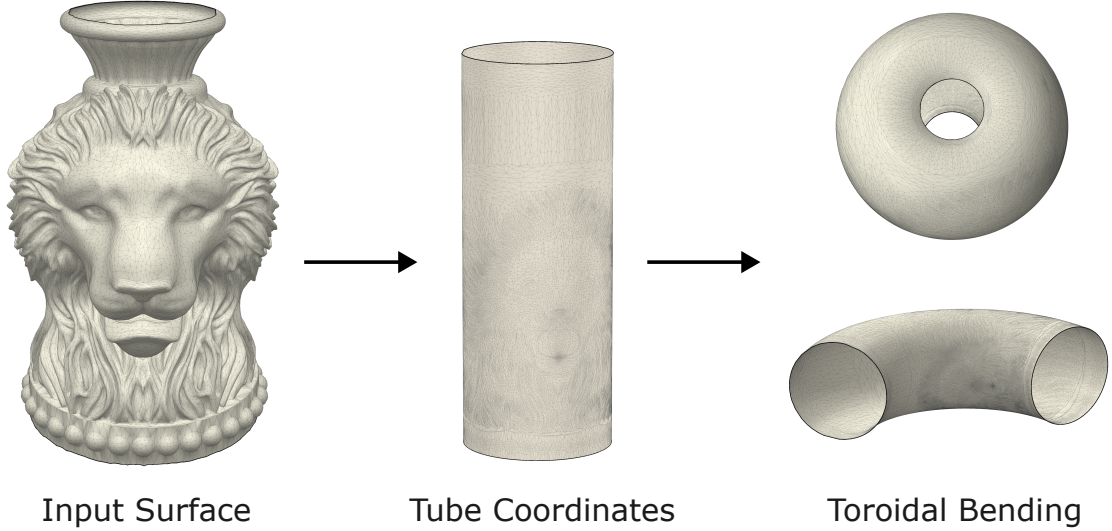}
    \caption{\textbf{Overview of the proposed conformal tubular parameterization and toroidal bending framework.} An input tube-like surface is conformally mapped to a tubular parameter domain and can further be conformally bent into toroidal geometries.}
    \label{fig:teaser}
\end{figure}

To address the above-mentioned issues, in this work we propose a flexible conformal parameterization framework for tube-like surfaces (Fig.~\ref{fig:teaser}). Building on the topology-aware viewpoint of cylindrical parameterization, our method maps the input surface to tubular or toroidally bent tube domains while explicitly controlling conformal distortion. While our framework builds on existing tools, including the linear Beltrami solver~\cite{Lui2013beltrami}, the rectangular conformal parameterization framework~\cite{Meng2016tempo}, and Laplacian smoothing~\cite{Taubin1995fairsurface}, its main contributions are threefold. First, we develop a tubular conformal parameterization method in which both the height and the angular shift of the parallelogram domain are optimized, so that the planar map can be conformally lifted to a cylinder with consistent periodic circumferential coordinates. Second, we introduce a localized quasi-conformal correction scheme, formulated on an annulus, that reduces seam-induced distortion while keeping the parameterization away from the seam unchanged. Third, we propose a free-boundary extension method that relaxes the boundary constraints for noisy or irregular boundaries. In addition, as an optional post-processing step, we derive two closed-form conformal toroidal bending maps for representing curved tube-like geometries and analyze their admissible radii. We validate the proposed methods on both synthetic examples and real medical vascular surfaces, showing that the framework provides controllable tubular and toroidal target domains with competitive angular distortion and improved adaptability to different geometries.

The rest of the paper is organized as follows. Section~\ref{sect:tube_conformal} presents the proposed conformal tubular parameterization framework, including the quasi-conformal preliminaries, the fixed-boundary tubular parameterization with localized seam correction and interior refinement, and the free-boundary extension pipeline. Section~\ref{sect:conformal_bending} introduces the conformal toroidal bending construction and the two associated bending modes. Section~\ref{sect:experiments} evaluates the proposed methods on synthetic and real vascular surfaces and reports ablation studies and computational cost. Section~\ref{sect:conclusion} concludes the paper and discusses possible future directions.

\section{Conformal Tubular Parameterization}\label{sect:tube_conformal}

In this section, we introduce our proposed conformal tubular parameterization framework.

\subsection{Quasi-conformal Map}\label{subsect:quasi_conformal}
We first briefly review the theoretical and computational aspects of quasi-conformal maps, which we use to control angular distortion in our parameterization problem. 

\paragraph{Theory of quasi-conformal maps}
Let $f:\mathbb{C}\to\mathbb{C}$ be a complex-valued map
\[
f(z)=f(x+iy)=u(x,y)+iv(x,y),
\]
where $u,v$ are real-valued $C^1$ functions. The map $f$ is conformal if and only if it satisfies the Cauchy--Riemann equations:
\begin{equation}\label{eq:cauchy_riemann}
    \begin{cases}
        u_x = v_y,\\
        u_y = -v_x.
    \end{cases}
\end{equation}

A quasi-conformal map generalizes conformality by satisfying the Beltrami equation
\begin{equation}\label{eq:beltrami}
    f_{\bar{z}}(z) = \mu_f(z) f_z(z),
\end{equation}
where
\[
f_{\bar{z}}=\frac{1}{2}(f_x+if_y),\qquad
f_z=\frac{1}{2}(f_x-if_y),
\]
and $\mu_f:\mathbb{C}\to\mathbb{C}$ is the Beltrami coefficient of $f$. When $\mu_f\equiv 0$, Eq.~\eqref{eq:beltrami} reduces to Eq.~\eqref{eq:cauchy_riemann}.

\begin{figure}
    \centering
    \includegraphics[width=\linewidth]{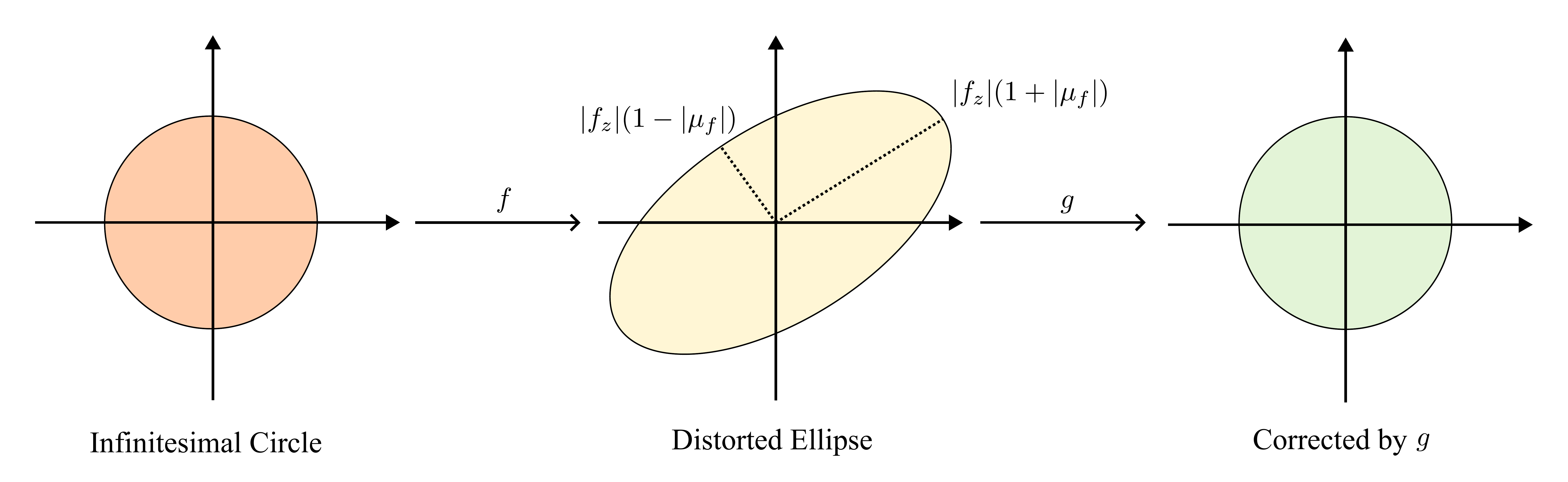}
    \caption{\textbf{Schematic illustration of distortion correction by quasi-conformal composition.} The map $f$ transforms an infinitesimal circle into an ellipse with principal stretches $|f_z|(1\pm|\mu_f|)$. Choosing a quasi-conformal map $g$ with $\mu_g \equiv\mu_{f^{-1}}$ cancels the anisotropy so that the composition satisfies $\mu_{g\circ f}=0$.}
    \label{fig:quasi_illu}
\end{figure}

The Beltrami coefficient quantifies local angular distortion. For a quasi-conformal map with $|\mu_f|<1$, the first fundamental form can be written as
\begin{equation*}
I = Df^\top Df
= |f_z|^2 Q^\top
\begin{pmatrix}
    (1+|\mu_f|)^2 & 0\\
    0 & (1-|\mu_f|)^2
\end{pmatrix}
Q
\end{equation*}
for some orthonormal matrix $Q$. Hence, an infinitesimal circle centered at $z$ is mapped to an infinitesimal ellipse centered at $f(z)$ (see Fig.~\ref{fig:quasi_illu}) with principal stretches
\[
\lambda_{\max}=|f_z(z)|(1+|\mu_f(z)|),\qquad
\lambda_{\min}=|f_z(z)|(1-|\mu_f(z)|).
\]

The Beltrami coefficient of a composition also has a closed form. For compatible quasi-conformal maps $f$ and $g$,
\begin{equation*}
    \mu_{g \circ f} = \frac{\mu_f+(\overline{f_z}/f_z)(\mu_g\circ f)}{1+(\overline{f_z}/f_z)\overline{\mu_f}(\mu_g\circ f)}.
\end{equation*}
If $\mu_g\equiv\mu_{f^{-1}}$, then $\mu_{g\circ f}=0$ and $g\circ f$ is conformal. This observation is the basis for distortion correction via quasi-conformal composition.

\paragraph{Computation of quasi-conformal maps}
By the Measurable Riemann Mapping Theorem~\cite{Ahlfors1960mapping}, a quasi-conformal map exists for any $\mu$ with $\|\mu\|_\infty<1$. Based on this result, Lui et al.~\cite{Lui2013beltrami} introduced the linear Beltrami solver (LBS) method to compute $f=u+iv$ for a prescribed $\mu_f$. From Eq.~\eqref{eq:beltrami},
\begin{equation*}
    \frac{f_{\bar{z}}}{f_z} = \frac{(u_x-v_y)+i(v_x+u_y)}{(u_x+v_y)+i(v_x-u_y)},
\end{equation*}
which is equivalent to
\begin{equation}\label{eq:beltrami_df}
    \begin{pmatrix}
        \alpha_1 & \alpha_2 \\
        \alpha_2 & \alpha_3
    \end{pmatrix}
    \begin{pmatrix}
        u_x & v_x \\
        u_y & v_y
    \end{pmatrix}
    =
    \begin{pmatrix}
        v_y & -u_y\\
        -v_x & u_x
    \end{pmatrix},
\end{equation}
where
\[
\alpha_1=\frac{(\text{Re}(\mu_f)-1)^2+\text{Im}(\mu_f)^2}{1-|\mu_f|^2},\quad
\alpha_2=-\frac{2 \ \text{Im}(\mu_f)}{1-|\mu_f|^2},\quad
\alpha_3=\frac{(\text{Re}(\mu_f)+1)^2+\text{Im}(\mu_f)^2}{1-|\mu_f|^2}.
\]
Using the differentiability of $u$ and $v$, Eq.~\eqref{eq:beltrami_df} leads to
\begin{equation}\label{eq:lbs_sys}
    \begin{cases}
        \nabla\cdot(A\nabla u)=0,\\
        \nabla\cdot(A\nabla v)=0,
    \end{cases}
\end{equation}
where $A = \begin{pmatrix}
            \alpha_1 & \alpha_2 \\
            \alpha_2 & \alpha_3
        \end{pmatrix}$.
In the discrete setting, Eq.~\eqref{eq:lbs_sys} reduces to two sparse symmetric positive-definite linear systems, which can be easily solved~\cite{Lui2013beltrami}. We denote the above linear Beltrami solver procedure by $f =\textbf{LBS}(\mu_f)$.

We note that Eq.~\eqref{eq:lbs_sys} follows from Eq.~\eqref{eq:beltrami_df} but is not equivalent to it. When boundary values are prescribed on the entire boundary, the solution of Eq.~\eqref{eq:lbs_sys} is the minimizer of the energy $\int (\nabla u^{\top}A\nabla u + \nabla v^{\top}A\nabla v)$ subject to the boundary constraints, and its Beltrami coefficient in general need not coincide with $\mu_f$.

\paragraph{Continuous and discrete views of conformal distortion}
Our computational framework combines the continuous viewpoint of quasi-conformal geometry with a discrete mesh-based realization. In the continuous setting, conformal distortion is characterized by the Beltrami coefficient, and the associated elliptic equations provide the basis for computing and correcting such distortion. On a triangulated surface, the mappings are discretized by finite elements, which yields the cotangent-type matrices used throughout, and the Beltrami coefficient is constant on each triangle and vanishes exactly when the triangle's angles are preserved. In this sense, the analytic quasi-conformal formulation and the discrete geometry viewpoint provide complementary descriptions of the same local distortion.

\subsection{Fixed-Boundary Conformal Tubular Parameterization}\label{sect:fix_para}
Given a connected, orientable, genus-zero triangulated open surface $M=(V,F)$ with exactly two boundary components $\partial M_0$ and $\partial M_1$, i.e., a surface topologically equivalent to an annulus (or an open cylinder), we construct a conformal tubular parameterization under circular boundary constraints. Specifically, the two boundaries in the final mapping result will maintain a fixed circular shape, while the positions of the vertices on the circular boundaries will be determined automatically to achieve conformality. The method comprises three stages: (i) initial tubular parameterization, (ii) cut seam correction, and (iii) interior refinement. Throughout this work, both the input surface $M$ and the target tube are two-dimensional surfaces embedded in $\mathbb{R}^3$.

\begin{figure}[t]
    \centering
    \includegraphics[width=\linewidth]{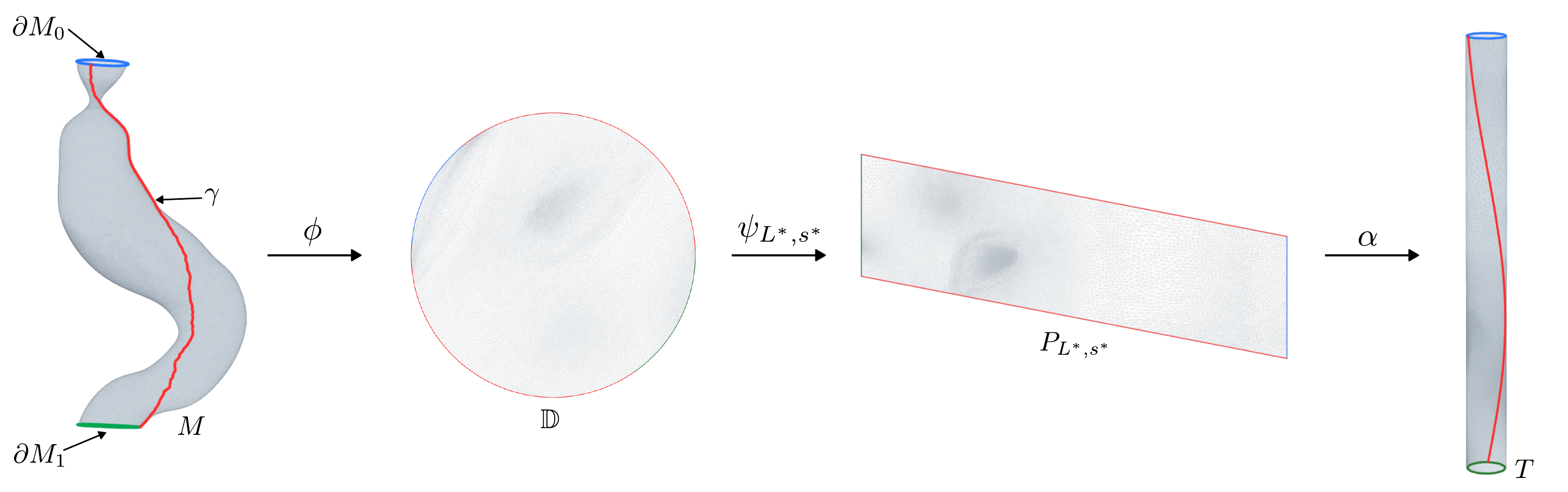}
    \caption{\textbf{Pipeline of the initial tubular parameterization.}  Starting from an open surface with two boundary loops (colored in blue and green), we cut the mesh along a shortest cut path (colored in red), compute a disk harmonic map $\phi$, transform it to a parallelogram via a quasi-conformal map $\psi_{L^*,s^*}$, and then lift it to a cylindrical surface via a conformal lifting map $\alpha$.}
    \label{fig:initial_tube}
\end{figure}

\paragraph{Initial tubular parameterization}
Our initial tubular parameterization builds on the rectangular conformal parameterization framework~\cite{Meng2016tempo} and is lifted by a conformal lifting map (see Fig.~\ref{fig:initial_tube}).

We first cut $M$ along a shortest boundary-to-boundary path $\gamma$ to convert the tube-topology mesh into a simply connected mesh $\widetilde{M}$. This step is essential because the subsequent disk and parallelogram parameterizations are posed on simply connected domains. 

Concretely, we construct a weighted graph $G=(V,E)$ from the mesh, where each edge $(v_i,v_j)\in E$ is assigned the Euclidean length $w_{ij}=\|v_i-v_j\|_2$. We then add two virtual source nodes, denoted by $s_0$ and $s_1$, where $s_0$ is connected to all vertices on $\partial M_0$ and $s_1$ is connected to all vertices on $\partial M_1$ using zero-weight edges. Running Dijkstra's algorithm~\cite{Dijkstra1959} from $s_0$ to $s_1$ yields a geodesic-like discrete shortest path, which we take as the cut seam $\gamma$.

After obtaining a simply connected mesh $\widetilde{M}$, we compute a disk harmonic map $\phi:\widetilde{M}\to\mathbb{D}$ by
\begin{equation*}
\begin{cases}
\Delta\phi=0,\\
\phi(\partial\widetilde{M})=\partial\mathbb{D},
\end{cases}
\end{equation*}
where the boundary values are set by arc-length parameterization. More concretely, let boundary vertices be $\{p_i\}_{i=1}^{n}$ in counterclockwise order, with edge lengths $\ell_i=\|p_{i+1}-p_i\|_2$ (cyclic indexing). We set
\[
\theta_1=0,\quad
\theta_i=2\pi\frac{\sum_{j=1}^{i-1}\ell_j}{\sum_{j=1}^{n}\ell_j},\quad i=2,\ldots,n,
\]
and map each $p_i$ to $(\cos\theta_i,\sin\theta_i)$. The Laplace operator $\Delta$ is discretized by the cotangent Laplacian $L_{\text{cotan}}$ with
\[
(L_{\text{cotan}})_{ij}=\begin{cases}
\dfrac{1}{2A_i}\big(\cot\alpha_{ij}+\cot\beta_{ij}\big), & j\in \mathcal{N}_V(i),\\
-\dfrac{1}{2A_i}\sum\limits_{k\in \mathcal{N}_V(i)}\big(\cot\alpha_{ik}+\cot\beta_{ik}\big), & j=i,\\
0, & \text{otherwise},
\end{cases}
\]
where $\mathcal{N}_V(i)$ denotes the one-ring vertex neighborhood of $v_i\in V$, $\alpha_{ij},\beta_{ij}$ are the two angles opposite to edge $(v_i,v_j)$ in the adjacent triangles and $A_i=\frac{1}{3}\sum_{T\in \mathcal{N}_F(i)}\operatorname{Area}(T)$ is the area weight associated with $v_i$, with $\mathcal{N}_F(i)$ denoting the one-ring face neighborhood of $v_i$.

We then compute a planar quasi-conformal map \(\psi_{L,s} : \mathbb{D} \to P_{L,s}\) with the prescribed target Beltrami coefficient $\mu_{\phi^{-1}}$, where \(P_{L,s}\) denotes a parallelogram parameter domain with longitudinal height $L$ and angular shift $s$ between its two boundary sides, and $\mu_{\phi^{-1}}$ is the Beltrami coefficient of the map $\phi^{-1}$:
\begin{equation*}
\psi_{L,s}=\textbf{LBS}(\mu_{\phi^{-1}}).
\end{equation*}
Denote by $p,p'$ the duplicated vertices at one endpoint of the cut path and by $q,q'$ those at the other endpoint on $\partial\widetilde{M}$. We impose corner constraints
\begin{equation*}
\psi_{L,s}(\phi(p))=(0,0),\ \ \psi_{L,s}(\phi(p'))=(0,2\pi), \ \ 
     \psi_{L,s}(\phi(q))=(L,s),\ \ \psi_{L,s}(\phi(q'))=(L,2\pi+s),
\end{equation*}
and enforce periodic consistency for corresponding vertices on the two duplicated cut boundaries (same $x$-coordinates). We determine the optimal height $L^*$ and angular shift $s^*$ by
\begin{equation*}
L^*,s^*=\mathop{\arg\min}\limits_{L>0, \ s} \|\mu_{\psi_{L,s}\circ\phi}\|^2,
\end{equation*}
which is solved using the derivative-free Powell method, initialized by a one-dimensional search over $L$ with $s=0$.
After obtaining the parallelogram parameterization $(z,u) \in P_{L^*,s^*}$, we map it to a standard tube $T = S^1 \times [0,L^*]$ via a conformal lifting map
\[
\alpha(z,u) = (\cos(u),\sin(u),z),
\]
and glue the cut path to reconstruct the tube topology. We denote the initial tubular parameterization map by $\eta = \alpha \circ (\psi_{L^*,s^*}) \circ \phi$.

\begin{figure}[t]
    \centering
    \includegraphics[width=\linewidth]{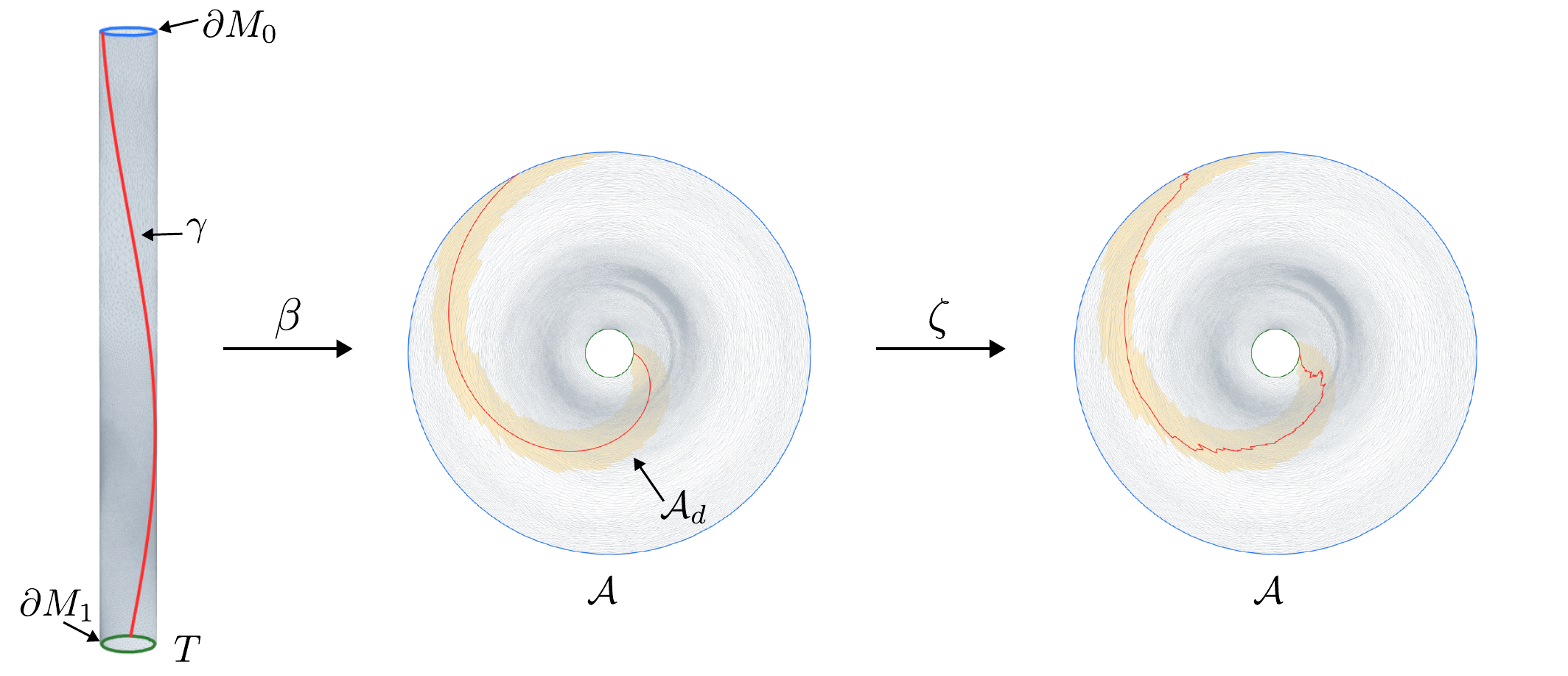}
    \caption{\textbf{Cut seam correction for the initial tubular parameterization.} The initial tubular parameterization $\eta$ is first transferred to the annulus $\mathcal{A}$ via $\beta$. A strip neighborhood (colored in yellow) of the cut seam is then corrected by a localized quasi-conformal map $\zeta$, while the rest of the annulus is kept fixed.}
    \label{fig:seam_correction}
\end{figure}

\paragraph{Cut seam correction}
$\eta$ gives a high-quality initial tubular parameterization. Nevertheless, residual angular distortion may still be concentrated near the cut seam, and an additional correction step is therefore needed (see Fig.~\ref{fig:seam_correction}). Since correcting distortion directly on the tube geometry is nontrivial, we transfer the problem to an annulus, where the correction can be formulated more conveniently~\cite{Choi2021multiconnect}.
Consider the conformal lifting map $\beta:T\to \mathcal{A}$ between the annulus $\mathcal{A} = \{w\in\mathbb{C}:1\le |w|\le e^{L^*}\}$ and the tube $T=S^1 \times [0,L^*]$, given by
\begin{equation*}
    \beta(x,y,z) = e^z\left( x+i y \right).
\end{equation*}
We then compute a localized angular-correction map on $\mathcal{A}$. Specifically, we only correct a narrow strip around the cut seam $\gamma$,
\[
\mathcal{A}_d = \beta \circ\eta(F_d), \quad F_d = \{ f \in F : \mathrm{dist}_G(v, \gamma) \le d \ \text{for all vertices } v \text{ of } f \},
\]
where $\mathrm{dist}_G(v,\gamma)$ is the shortest-path distance from $v$ to $\gamma$ on the mesh graph, with each edge $(v_k, v_l)$ weighted by its length $\|\eta(v_k) - \eta(v_l)\|_2$ in the initial tube coordinates. Since the tube has unit radius, $d$ is measured relative to the tube radius and the strip has a uniform width along the seam.

This localization is important because applying a global correction directly to an annular shape is highly unstable and may introduce substantial additional angular distortion. Let
\[
\widetilde{\mu}(w)=
\begin{cases}
\mu_{\eta^{-1}\circ\beta^{-1}}(w), & w\in\mathcal{A}_d,\\
0, & w\in\mathcal{A}\setminus\mathcal{A}_d.
\end{cases}
\]
On $\mathcal{A}_d$, we then solve
\begin{equation*}
    \zeta_d=\textbf{LBS}(\widetilde{\mu}),
\end{equation*}
with the boundary condition $\zeta_d(w)=w$ for all $w\in\partial\mathcal{A}_d$. The global correction map is defined by
\begin{equation*}
    \zeta(w)=
    \begin{cases}
        \zeta_d(w), & w\in\mathcal{A}_d,\\
        w, & w\in\mathcal{A}\setminus\mathcal{A}_d.
    \end{cases}
\end{equation*}
By construction, $\zeta$ redistributes distortion only near the seam while leaving the rest of the annulus unchanged. 
Combining these mappings together, we obtain the seam-corrected parameterization
\[\eta_{\text{seam}} = \zeta \circ \beta \circ \eta.\]

In general, a prescribed Beltrami coefficient and identity data on the entire boundary $\partial\mathcal{A}_d$ are not guaranteed to be compatible, so the Beltrami coefficient of $\zeta_d$ need not coincide with $\widetilde{\mu}$. We nevertheless impose this condition so that $\zeta$ is continuous across $\partial\mathcal{A}_d$ and leaves $\mathcal{A}\setminus\mathcal{A}_d$ unchanged. The subsequent interior refinement is intended to further reduce the residual distortion.

\begin{figure}[t]
    \centering
    \includegraphics[width=\linewidth]{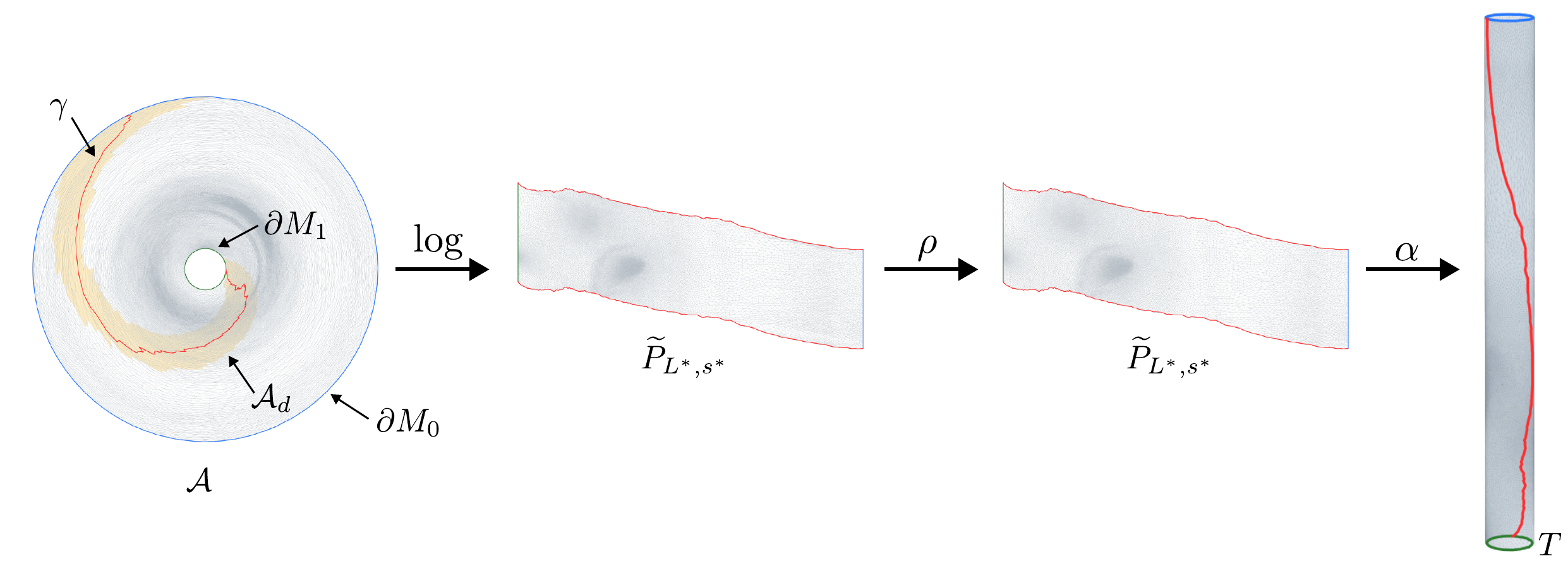}
    \caption{\textbf{Interior refinement after cut seam correction.} The seam-corrected annulus is cut open along the corrected seam and lifted to a parallelogram-like domain by choosing the logarithmic branch induced by the seam. A final quasi-conformal map $\rho$ is then computed with the boundary fixed, allowing only the interior vertices to be adjusted. The refined parallelogram parameterization is finally wrapped back to the tube by $\alpha$.}
    \label{fig:interior_refinement}
\end{figure}

\paragraph{Interior refinement}
Although the cut seam correction step reduces the distortion around the seam, the interior of the parameter domain may still contain residual distortion inherited from the initial map. To further improve the interior while preserving the corrected seam and boundary alignment, we cut the annulus $\mathcal{A}$ open along the corrected seam to obtain the simply connected domain $\widetilde{\mathcal{A}}$ and conformally map the corrected tube coordinates back onto a parallelogram-like domain 
\[\widetilde{P}_{L^*,s^*} = \log(\widetilde{\mathcal{A}})\]
by the logarithmic map with a branch cut induced by $\gamma$ (see Fig.~\ref{fig:interior_refinement}). Here, $\widetilde{P}_{L^*,s^*}$ is a parallelogram-like domain bounded by the segments $z=0$ and $z=L^*$ and by the images of the two sides of the corrected seam, which differ by a translation of $2\pi$ in $u$ but are in general no longer straight.

We then compute a final quasi-conformal correction on the cut-open domain by solving
\[
  \rho=\textbf{LBS}(\mu_{\eta_{\text{seam}}^{-1} \circ \log^{-1}}),
\]
with the whole boundary shape of $\widetilde{P}_{L^*,s^*}$ fixed. As discussed above, $\rho$ therefore does not necessarily realize the target Beltrami coefficient exactly. The final tubular parameterization is obtained by wrapping the corrected parallelogram coordinates back to the tube via $\alpha$:
\[
\Phi_{\mathrm{fix}} = \alpha \circ \rho \circ \log \circ \eta_{\text{seam}}.
\]

The complete fixed-boundary conformal tubular parameterization pipeline is summarized as
\[
M\xrightarrow{\text{cut}}\widetilde{M}
\xrightarrow{\phi}\mathbb{D}
\xrightarrow{\psi_{L^*,s^*}}P_{L^*,s^*}
\xrightarrow{\alpha}T
\xrightarrow{\beta}\mathcal{A}
\xrightarrow{\zeta}\mathcal{A}
\xrightarrow{\text{cut}}\widetilde{\mathcal{A}}
\xrightarrow{\log}\widetilde{P}_{L^*,s^*}
\xrightarrow{\rho}\widetilde{P}_{L^*,s^*}
\xrightarrow{\alpha}T,
\]
and Algorithm~\ref{alg:fixed_tube_param} summarizes the implementation details.

\begin{algorithm}
\caption{Fixed-boundary conformal tubular parameterization}
\label{alg:fixed_tube_param}
\begin{algorithmic}[1]
\STATE \textbf{Input:} Tube-topology mesh $M=(V,F)$ with boundary loops $\partial M_0,\partial M_1$
\STATE \textbf{Output:} Fixed-boundary conformal tubular parameterization $\Phi_{\mathrm{fix}}:M\to T$
\STATE Build weighted graph $G$ and find seam $\gamma$ between virtual sources linked to $\partial M_0$ and $\partial M_1$ by Dijkstra's algorithm.
\STATE Cut $M$ along $\gamma$ to obtain simply connected mesh $\widetilde{M}$.
\STATE Compute disk harmonic map $\phi:\widetilde{M}\to\mathbb{D}$ with arc-length boundary constraints.
\STATE Solve $\psi_{L,s}=\textbf{LBS}(\mu_{\phi^{-1}})$ under corner/periodic constraints and optimize $L^*,s^*=\arg\min_{L>0,s}\|\mu_{\psi_{L,s}\circ\phi}\|^2$.
\STATE Lift the planar result to a tube by $\eta=\alpha\circ\psi_{L^*,s^*}\circ\phi$.
\STATE Map the tube to an annulus via $\beta$, solve a localized quasi-conformal correction on strip $\mathcal{A}_d$ with fixed strip boundary and obtain the seam-corrected map $\eta_{\text{seam}} = \zeta \circ \beta \circ\eta$.
\STATE Cut $\eta_{\text{seam}}(M)$ open along the corrected seam and map it to a parallelogram-like domain $\widetilde{P}_{L^*,s^*}$ via the logarithmic map.
\STATE Solve a final LBS problem $\rho=\textbf{LBS}(\mu_{\eta_{\text{seam}}^{-1} \circ \log^{-1}})$ on $\widetilde{P}_{L^*,s^*}$ with the new boundary shape fixed and lift it to a tube with $\alpha$.
\STATE Return  $\Phi_{\mathrm{fix}}=\alpha \circ \rho \circ \log \circ \eta_{\text{seam}}$.
\end{algorithmic}
\end{algorithm}

\subsection{Free-Boundary Conformal Tubular Parameterization}
\begin{figure}[t!]
    \centering
    \includegraphics[width=\linewidth]{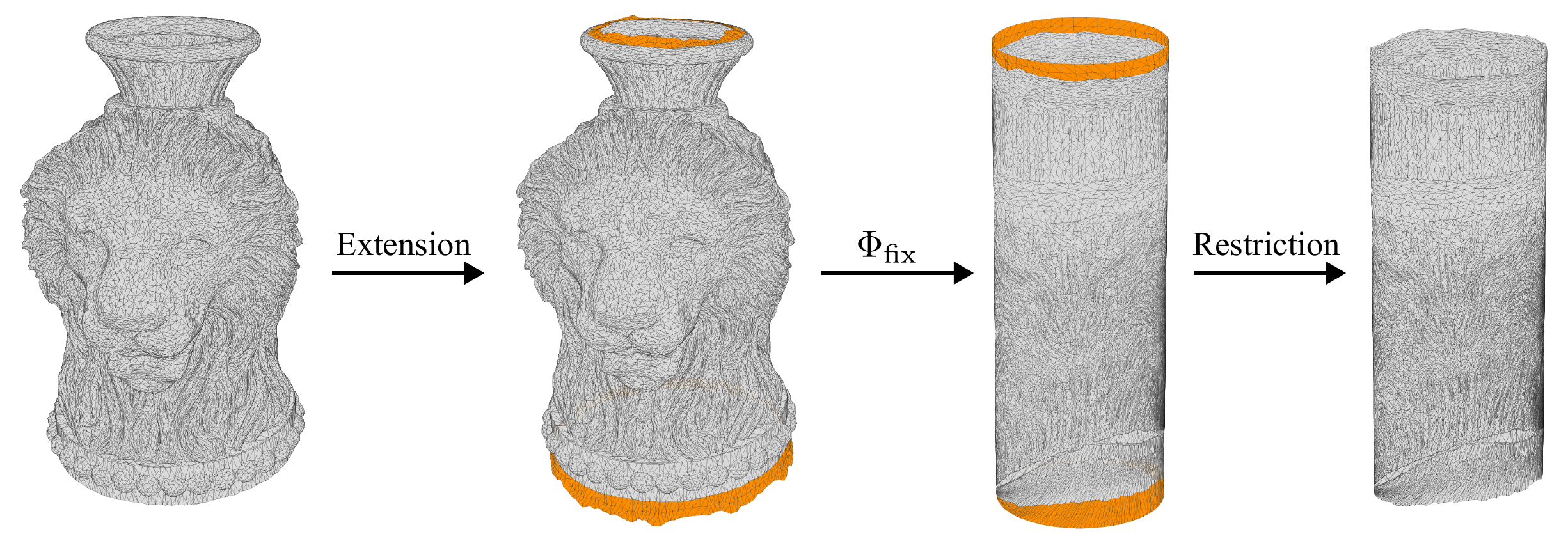}
    \caption{\textbf{Pipeline of the free-boundary conformal tubular parameterization.} Boundary loops are extended and smoothed to form an augmented mesh $M_{\mathrm{ext}}$ (extended part is colored in orange). The fixed-boundary parameterization $\Phi_{\mathrm{fix}}$ is then applied to $M_{\mathrm{ext}}$, and the result is finally restricted back to the original mesh $M$.}
    \label{fig:free_pipeline}
\end{figure}

Directly imposing circular boundary constraints on the tubular parameterization may introduce boundary bias, particularly when the input boundaries are noisy or highly irregular. Inspired by virtual-boundary and scaffold-mesh techniques from the early parameterization literature~\cite{Zayer2005boundaryfree}, we consider a free-boundary conformal tubular parameterization pipeline consisting of three steps (see Fig.~\ref{fig:free_pipeline}): (i) raw extension of boundary loops, (ii)~cycle-Laplacian smoothing, and (iii) restriction of the result back to the original surface.

\begin{figure}[t]
    \centering
      \includegraphics[width=0.8\linewidth]{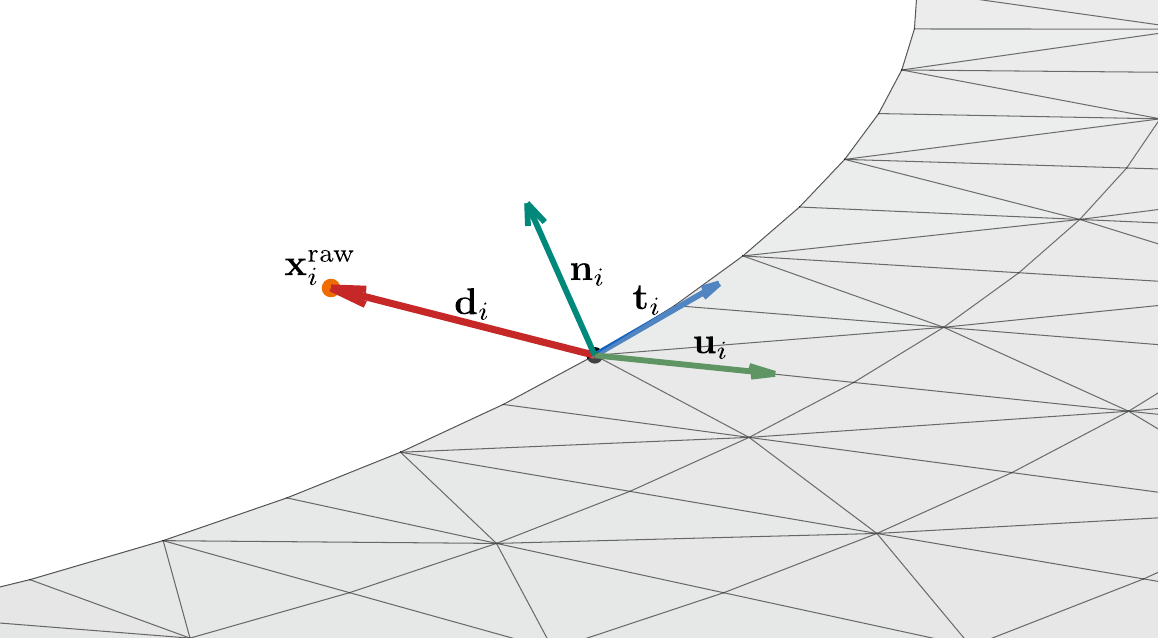}
    \caption{\textbf{Raw boundary extension on a boundary loop.} At each boundary vertex, an outward direction $\mathbf d_i$ is estimated from the local tangent $\mathbf t_i$, the normal $\mathbf n_i$ and the inward direction $\mathbf u_i$.}
    \label{fig:raw_extension}
\end{figure}

\paragraph{Raw extension of boundary loops}
Let the two boundary loops be $\partial M_0$ and $\partial M_1$. For each boundary loop, we iteratively append $K$ triangle layers. At each iteration, we denote the current boundary ring by $\{\mathbf p_i\}_{i=1}^{m}$ (cyclic indexing) and define
\[
\mathring{\mathcal V}_i = \{q\in \mathcal N_V(i) : q\ \text{is not on the current ring}\}.
\]
We first compute the boundary tangent $\mathbf t_i$ and an approximate inward direction $\mathbf u_i$ (see Fig.~\ref{fig:raw_extension}):
\[
\mathbf{t}_i= \frac{\mathbf{p}_{i+1}-\mathbf{p}_{i-1}}{\|\mathbf{p}_{i+1}-\mathbf{p}_{i-1}\|_2}, \qquad
\mathbf u_i= \frac{\sum_{q\in \mathring{\mathcal V}_i}}(q-\mathbf p_i){\left\|\sum_{q\in \mathring{\mathcal V}_i} (q-\mathbf p_i)\right\|_2}.
\]
A candidate lateral extension direction is given by
\[
\widetilde{\mathbf b}_i= \frac{\mathbf t_i\times\mathbf n_i}{\|\mathbf t_i\times\mathbf n_i\|_2},
\]
where $\mathbf n_i$ is an area-weighted incident face normal vector. Since $\widetilde{\mathbf b}_i$ may point inward or outward depending on local orientation, we correct its sign by
\[
\mathbf b_i=
\begin{cases}
-\widetilde{\mathbf b}_i, & \text{if }\widetilde{\mathbf b}_i\cdot\mathbf u_i>0,\\
\widetilde{\mathbf b}_i, & \text{otherwise}.
\end{cases}
\]
We then combine the lateral and normal components to improve robustness:
\[
\mathbf d_i= \frac{(1-\tau)\mathbf b_i+\tau\mathbf n_i}{\|(1-\tau)\mathbf b_i+\tau\mathbf n_i\|_2},
\]
where $\tau \in[0,1]$ is the combination coefficient.
The point-wise step length is estimated from the interior spacing
\[
s_i=\frac{1}{| \mathring{\mathcal V}_i|}\sum_{q\in\mathring{\mathcal V}_i}\|q-\mathbf p_i\|_2.
\]
Finally, the raw extension is given by
\begin{equation*}
    \mathbf x_i^{\mathrm{raw}}=\mathbf p_i+ \operatorname{mean}(s_i)\mathbf d_i,
\end{equation*}
where we adopt the mean of all point-wise step lengths to reduce the influence of degenerate triangles near the boundary loops.

\begin{figure}[t]
    \centering
    \includegraphics[width=\linewidth]{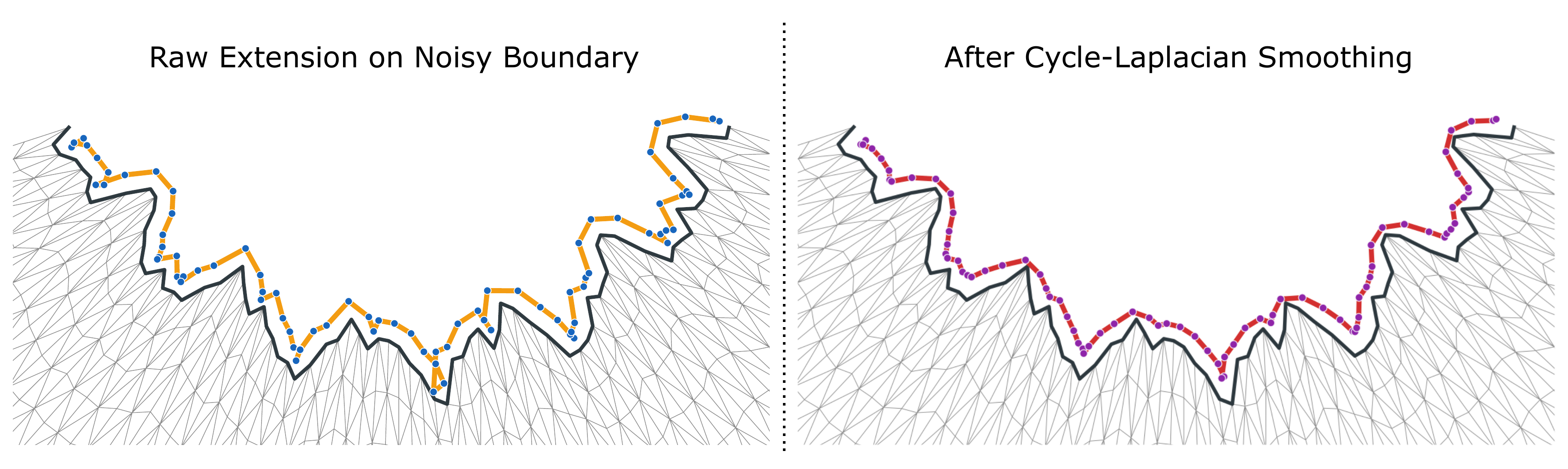}
    \caption{\textbf{Cycle-Laplacian smoothing of the raw extension.} The smoothed ring balances data fidelity and cycle smoothness, thereby suppressing high-frequency zigzags.}
    \label{fig:laplacian_smooth}
\end{figure}

\paragraph{Cycle-Laplacian smoothing} 
The raw extension is determined from local normals and local spacing and is therefore sensitive to uneven sampling and geometric noise. High-frequency zigzags may appear, leading to poorly shaped triangles and reduced stability in subsequent quasi-conformal correction on the original surface. We therefore formulate an energy minimization problem based on Laplacian smoothing~\cite{Taubin1995fairsurface} on the raw extension while keeping it close to the intended outward displacement (see Fig.~\ref{fig:laplacian_smooth} for an illustration). At the discrete graph level, this can be interpreted as an implicit Laplacian smoothing step with data attachment: high-frequency zigzags are attenuated, while the new boundary remains close to the intended outward offset~\cite{Kazhdan2012MeanCurvature,Desbrun1999ImplicitFairing}. This first-order graph regularization is sufficient for stabilizing the extension strip; in contrast, higher-order fairing flows such as Willmore flow optimize global fairness and may introduce shrinkage or displacement away from the feasible outward direction, which is not desirable for this auxiliary boundary-extension step.

More specifically, we find the smoothed extension $\{\mathbf x_i\}_{i=1}^{m}$ (cyclic indexing) by minimizing the energy \vspace{-5mm}
\begin{equation*}
E_{\text{smooth}}= \sum_{i=1}^{m}\|\mathbf x_i-\mathbf x_i^{\mathrm{raw}}\|_2^2
+ \omega\sum_{i=1}^{m}\|\mathbf x_{i+1}-\mathbf x_i\|_2^2,
\end{equation*}
where $\omega$ is a weighting coefficient.
This is a convex quadratic problem, and hence imposing the first-order optimality condition $\nabla E_{\text{smooth}} = 0$ yields a sparse symmetric linear system
\begin{equation*}
    \left(I+\omega L_{\mathrm{cycle}}\right)\mathbf x = \mathbf x^{\mathrm{raw}},
\end{equation*}
where the $i$-th row of $\mathbf x \in \mathbb{R}^{m\times 3}$ is $\mathbf x_i$ (similarly for $\mathbf x^{\mathrm{raw}}$) and $L_{\mathrm{cycle}}$ is the cycle-graph Laplacian matrix given by
\[
(L_{\mathrm{cycle}})_{ij}=\begin{cases}
2, & i=j,\\
-1, & j=i-1 \text{ or } j=i+1 \text{ in cyclic indexing},\\
0, & \text{otherwise}.
\end{cases}
\]

Here, $L_{\mathrm{cycle}}$ is used as a combinatorial graph regularizer on the ordered boundary samples, rather than as a finite-element discretization of a curve flow. Hence, we do not introduce a lumped or consistent mass matrix and use the identity data term to attach each raw extension vertex equally to its prescribed outward displacement. Since the eigenvalues of the cycle-graph Laplacian satisfy $0\leq\lambda(L_{\mathrm{cycle}})\leq 4$, the matrix $I+\omega L_{\mathrm{cycle}}$ is symmetric positive definite with $\kappa(I+\omega L_{\mathrm{cycle}})\leq 1+4\omega$. Thus the implicit solve is unconditionally stable in the usual backward-Euler sense; the parameter $\omega$ controls only the amount of smoothing.

\paragraph{Restricting the result back to the original surface}
After extending both boundaries, we obtain an augmented mesh $M_{\mathrm{ext}}$. We then apply the fixed-boundary pipeline in Section~\ref{sect:fix_para} to compute $\Phi_{\mathrm{ext}}: M_{\mathrm{ext}}\to T$. The final free-boundary parameterization on the original mesh is defined by
\[
\Phi=\Phi_{\mathrm{ext}}\big|_M.
\]
Since explicit boundary constraints are imposed only on the artificial outer rings, the original boundaries are no longer forced onto prescribed circles, thereby reducing boundary bias while maintaining low angular distortion.

Algorithm~\ref{alg:free_tube_param} summarizes the procedure of free-boundary parameterization.

\begin{algorithm}
\caption{Free-boundary conformal tubular parameterization}
\label{alg:free_tube_param}
\begin{algorithmic}[1]
\STATE \textbf{Input:} Mesh $M$ with boundary loops $\partial M_0,\partial M_1$, number of extension layers $K$, combination coefficient $\tau$, smoothing weight $\omega$
\STATE \textbf{Output:} Free-boundary conformal tubular parameterization $\Phi:M\to T$
\FOR{each boundary loop $\partial M_b$, $b\in\{0,1\}$}
    \FOR{$k=1,\ldots,K$}
        \STATE Estimate local directions $(\mathbf t_i,\mathbf u_i,\mathbf n_i)$, compute the oriented lateral direction $\mathbf b_i$, and set the outward direction as $\mathbf d_i=\frac{(1-\tau)\mathbf b_i+\tau\mathbf n_i}{\|(1-\tau)\mathbf b_i+\tau\mathbf n_i\|_2},$ for each vertex on the current boundary ring.
        \STATE Estimate the interior spacing $s_i$ of each boundary vertex and compute raw ring points $\mathbf x_i^{\mathrm{raw}}=\mathbf p_i+ \operatorname{mean}(s_i)\mathbf d_i$.
        \STATE Solve the cycle-Laplacian smoothing system $(I+\omega L_{\mathrm{cycle}})\mathbf x=\mathbf x^{\mathrm{raw}}$ to obtain the smoothed ring $\{\mathbf x_i\}$.
        \STATE Stitch the smoothed ring to the mesh and update it as the current boundary loop.
    \ENDFOR
\ENDFOR
\STATE Obtain the augmented mesh $M_{\mathrm{ext}}$ after extending both boundary loops.
\STATE Compute $\Phi_{\mathrm{ext}}:M_{\mathrm{ext}}\to T$ using Algorithm~\ref{alg:fixed_tube_param}.
\STATE Return $\Phi=\Phi_{\mathrm{ext}}\big|_M$.
\end{algorithmic}
\end{algorithm}

\section{Conformal Tube Bending onto Toroidal Geometry}\label{sect:conformal_bending}
In the previous section, we established a methodology for obtaining a conformal tubular parameterization of an open surface with two boundary components. While the straight tube provides a natural parameter domain for tube-like surfaces, it may not fully reflect the global geometry of highly curved, nearly closed, or ring-like tubular shapes. We therefore introduce conformal toroidal bending as an optional post-processing step that provides additional flexibility in the choice of the target geometry. In particular, the bending maps preserve conformality and the longitudinal and periodic circumferential coordinate structure of the tubular parameterization, while the major radius serves as an additional degree of freedom that allows the target domain to accommodate global curvature and thereby reduce area distortion. We derive two such maps: a minor-mode map, which converges to the straight tube as the major radius increases and thus serves as a general-purpose target, and a major-mode map, which is intended for applications requiring a full wrapping along the major circle.

Specifically, given a tubular parameterization $\Phi:M\to T$, we construct a conformal bending map $\mathcal{B}$ from the straight tube $T=S^1 \times [0,L]$ into the torus $\mathbb{T}=S^1\times S^1$ embedded in $\mathbb{R}^3$. The final target map is given by
\[
\Psi=\mathcal{B}\circ \Phi.
\]

We emphasize that, since $\mathcal{B}$ is conformal, $\Psi$ and $\Phi$ induce the same conformal structure and the same tube coordinates $(u,z)$; the two parameterizations differ only by a conformal factor, i.e., in the extrinsic geometry of the target domain and the resulting area distribution, rather than in the underlying conformal parameterization.

\paragraph{Bending formulation in tube coordinates}
Let a torus with major radius $R>1$ and minor radius $r=1$ be parameterized by
\begin{equation*}
    X(\theta,\phi)=\big((R+\cos\theta)\cos\phi,\,(R+\cos\theta)\sin\phi,\,\sin\theta\big).
\end{equation*}
The tangent vectors of $X$ are
\[
X_\theta=(-\sin\theta\cos\phi,\,-\sin\theta\sin\phi,\,\cos\theta),\ \ X_\phi=(-(R+\cos\theta)\sin\phi,\,(R+\cos\theta)\cos\phi,\,0).
\]
The choice $r=1$ is without loss of generality: for a general minor radius $r$, the conformality condition below depends only on the ratio $R/r$, and the resulting bending map differs from the case $r=1$ only by a uniform scaling, which does not affect conformality or the normalized distortion measures. We take $r=1$ to match the unit cross-section of the tube $T$.

\begin{figure}[t]
    \centering
    \includegraphics[width=\linewidth]{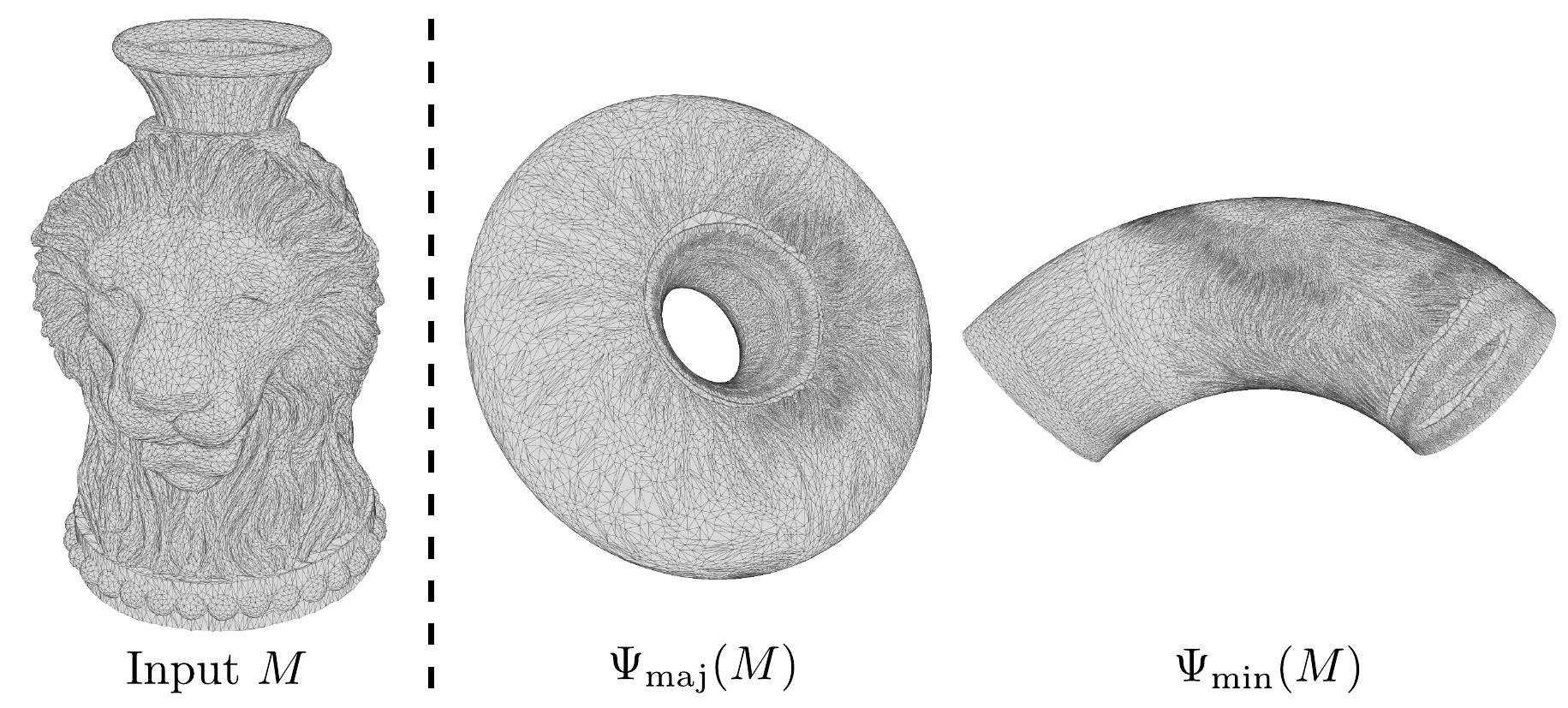}
    \caption{\textbf{Examples of conformal toroidal bending using the two different bending maps.} The results are obtained by full wrapping along the major circle $\Psi_{\mathrm{maj}}$ and full wrapping along the minor circle $\Psi_{\mathrm{min}}$, respectively.}
    \label{fig:tube_bent}
\end{figure}

Given tube coordinates $(u,z)$, we consider a general differentiable bending map
\[
B(u,z)=X\big(\theta(u,z),\phi(u,z)\big).
\]
Differentiating $B$ gives
\[
B_u=X_\theta \theta_u+X_\phi \phi_u,\qquad B_z=X_\theta \theta_z+X_\phi \phi_z,
\]
and hence the first fundamental form is
\[
ds^2 = E\,du^2+2F\,du\,dz + G\,dz^2,
\]
where
\[
E = \theta_u^2+(R+\cos\theta)^2\phi_u^2,\ \ F = \theta_u\theta_z+(R+\cos\theta)^2\phi_u\phi_z,\ \ G = \theta_z^2+(R+\cos\theta)^2\phi_z^2.
\]
To construct a conformal bending map, we impose $E=G$ and $F=0$~\cite{Gu2003GlobalConformal}. A tractable choice is to assume separated variables~\cite{Guenther2020superfluid}, e.g., $\theta=\theta(x)$ and $\phi=\phi(y)$, so that $F=0$ automatically. The conformality condition then reduces to
\begin{equation}\label{eq:conformal_cond}
    \left(\frac{1}{R+\cos\theta} \frac{d\theta}{dx}\right)^2 = \left(\frac{d\phi}{dy}\right)^2.
\end{equation}
Since $R+\cos\theta > 0$ and the two sides of Eq.~\eqref{eq:conformal_cond} depend on different variables, both sides must equal a constant $c^2$ with $c>0$:
\[
\left(\frac{1}{R+\cos\theta} \frac{d\theta}{dx}\right)^2 = \left(\frac{d\phi}{dy}\right)^2 = c^2.
\]
Without loss of generality (up to orientation), we take
\[
\frac{1}{R+\cos\theta} \frac{d\theta}{dx}= \frac{d\phi}{dy} = c.
\]
Integrating with initial conditions $\theta(0)=0$ and $\phi(0)=0$ yields
\begin{equation*}
    \begin{cases}
         \theta(x) = 2 \arctan\left( \sqrt{\frac{R+1}{R-1}} \tan \left( \frac{\sqrt{R^2-1}}{2} cx \right) \right), \\
         \phi(y) = cy.
    \end{cases}
\end{equation*}
We then obtain two bending maps by assigning $x,y$ to $u,z$ in two different ways (see Fig.~\ref{fig:tube_bent}), as detailed below.

\paragraph{Full wrapping along the major circle}
We assign $x=z$ and $y=u$ so that
\begin{equation}\label{eq:major_param}
\begin{cases}
\theta_{\mathrm{maj}}(z)=2\arctan\left(\sqrt{\frac{R+1}{R-1}}\tan\left(\frac{\sqrt{R^2-1}}{2}c_{\mathrm{maj}}z\right)\right), \\
\phi_{\mathrm{maj}}(u)=c_{\mathrm{maj}}u,
\end{cases}
\end{equation}
where $c_{\mathrm{maj}} > 0$ is a constant associated with the major revolution. This assignment realizes a full wrapping along the major circle.

Since $u\in[0,2\pi)$ is periodic, one complete wrapping around the major circle requires
\[
\phi_{\mathrm{maj}}(2\pi)-\phi_{\mathrm{maj}}(0)=2\pi.
\]
Hence $c_{\mathrm{maj}}=1$, and Eq.~\eqref{eq:major_param} reduces to
\[
\begin{cases}
\theta_{\mathrm{maj}}(z)=2\arctan\left(\sqrt{\frac{R+1}{R-1}}\tan\left(\frac{\sqrt{R^2-1}}{2}z\right)\right), \\
\phi_{\mathrm{maj}}(u)=u.
\end{cases}
\]
The corresponding conformal bending map is
\begin{equation*}
\mathcal B_{\mathrm{maj}}(u,z)=\Big((R+\cos\theta_{\mathrm{maj}}(z))\cos u,\,(R+\cos\theta_{\mathrm{maj}}(z))\sin u,\,\sin\theta_{\mathrm{maj}}(z)\Big),
\end{equation*}
with conformal factor
\[
\lambda_{\mathrm{maj}}(z)=R+\cos\theta_{\mathrm{maj}}(z),
\]
where the bending metric is $ds^2=\lambda_{\mathrm{maj}}(z)^2(du^2+dz^2)$.
Since $R>1$, we have $\theta_{\mathrm{maj}}'(z)=R+\cos\theta_{\mathrm{maj}}(z)>0$ and hence the map is locally orientation-preserving without axial fold-over.
Define
\[
\Delta\theta_{\mathrm{maj}}=\theta_{\mathrm{maj}}(\max(z)) - \theta_{\mathrm{maj}}(\min(z)),
\]
which measures the meridional extent. To avoid wrap-around self-overlap in the minor direction, we impose $\Delta\theta_{\mathrm{maj}}<2\pi$. Set $\Delta z = \max(z)-\min(z)$. This condition can be enforced by choosing \vspace{-5mm}
\begin{equation}\label{eq:maj_no_overlap}
R < R_{\mathrm{maj}}^{+}:=\sqrt{1+\left(\frac{2\pi}{\Delta z}\right)^2}.
\end{equation}
The resulting parameterization is
\[
\Psi_{\mathrm{maj}}=\mathcal B_{\mathrm{maj}}\circ\Phi.
\]

\paragraph{Full wrapping along the minor circle}
Similarly, we assign $x=u$ and $y=z$, giving
\begin{equation}\label{eq:minor_param}
\begin{cases}
\theta_{\mathrm{min}}(u)=2\arctan\left(\sqrt{\frac{R+1}{R-1}}\tan\left(\frac{\sqrt{R^2-1}}{2}c_{\mathrm{min}}u\right)\right), \\
    \phi_{\mathrm{min}}(z)=c_{\mathrm{min}}z,
\end{cases}
\end{equation}
where $c_{\mathrm{min}} > 0$ is a constant associated with the minor revolution. 

To enforce one complete cycle along the minor circle, we require
\[
\theta_{\mathrm{min}}(2\pi)-\theta_{\mathrm{min}}(0)=2\pi.
\]
Using $\theta'(u)=c_{\mathrm{min}}(R+\cos\theta)$, we obtain
\[
2\pi c_{\mathrm{min}}=\int_{0}^{2\pi}\frac{d\theta}{R+\cos\theta}=\frac{2\pi}{\sqrt{R^2-1}} \implies c_{\mathrm{min}}=\frac{1}{\sqrt{R^2-1}}.
\]
Hence Eq.~\eqref{eq:minor_param} becomes
\[
\begin{cases}
    \theta_{\mathrm{min}}(u)=2\arctan\left(\sqrt{\frac{R+1}{R-1}}\tan\left(\frac{u}{2}\right)\right),\\
    \phi_{\mathrm{min}}(z)=\frac{z}{\sqrt{R^2-1}}.
\end{cases}
\]
Therefore, the conformal bending map is
\begin{equation*}
\mathcal B_{\mathrm{min}}(u,z)=\Big((R+\cos\theta_{\mathrm{min}}(u))\cos\phi_{\mathrm{min}}(z),\,(R+\cos\theta_{\mathrm{min}}(u))\sin\phi_{\mathrm{min}}(z),\,\sin\theta_{\mathrm{min}}(u)\Big),
\end{equation*}
with conformal factor
\[
\lambda_{\mathrm{min}}(u)=\frac{R+\cos\theta_{\mathrm{min}}(u)}{\sqrt{R^2-1}}.
\]
Since $\theta'_{\mathrm{min}}(u)=\frac{1}{\sqrt{R^2-1}}(R+\cos\theta_{\mathrm{min}}(u))>0$, the map is locally orientation-preserving.
Moreover,
\[
\Delta\phi_{\mathrm{min}}=\phi_{\mathrm{min}}(\max(z))-\phi_{\mathrm{min}}(\min(z))=\frac{\Delta z}{\sqrt{R^2-1}}
\]
determines the major-direction coverage. To avoid wrap-around self-overlap in the major direction, we impose $\Delta\phi_{\mathrm{min}}<2\pi$. This condition can be enforced by choosing
\begin{equation}\label{eq:min_no_overlap}
R > R_{\min}^{-}:=\sqrt{1+\left(\frac{\Delta z}{2\pi}\right)^2}.
\end{equation}
The resulting parameterization is
\[
\Psi_{\mathrm{min}}=\mathcal B_{\mathrm{min}}\circ\Phi.
\]
Algorithm~\ref{alg:tube_torus_bending} summarizes the conformal tube bending procedures.

\begin{algorithm}
\caption{Conformal tube bending}
\label{alg:tube_torus_bending}
\begin{algorithmic}[1]
\STATE \textbf{Input:} Tubular parameterization $\Phi:M\to T$, major radius $R>1$, mode $m\in\{\mathrm{major},\mathrm{minor}\}$
\STATE \textbf{Output:} Toroidal parameterization $\Psi:M\to\mathbb T$
\STATE Compute $z_{\min}=\min(z)$ and $z_{\max}=\max(z)$ on $\Phi(M)$ and set $\Delta z=z_{\max}-z_{\min}$.
\STATE Normalize $\hat{z} = z - \min(z)$.
\IF{$m=\mathrm{major}$}
    \STATE Choose $R<\sqrt{1+\left(\dfrac{2\pi}{\Delta z}\right)^2}$ to avoid minor-direction overlap.
    \STATE Set $\theta(\hat{z})=2\arctan\!\left(\sqrt{\dfrac{R+1}{R-1}}\tan\!\left(\dfrac{\sqrt{R^2-1}}{2}\hat{z}\right)\right)$ and $\phi(u)=u$.
    \STATE Define $\mathcal B_{\mathrm{maj}}(u,z)=X(\theta(\hat{z}),\phi(u))$ and set $\Psi=\mathcal B_{\mathrm{maj}} \circ\Phi$.
\ELSE
    \STATE Choose $R>\sqrt{1+\left(\dfrac{\Delta z}{2\pi}\right)^2}$ to avoid major-direction overlap.
    \STATE Set $\theta(u)=2\arctan\!\left(\sqrt{\dfrac{R+1}{R-1}}\tan\!\left(\dfrac{u}{2}\right)\right)$ and $\phi(\hat{z})=\dfrac{\hat{z}}{\sqrt{R^2-1}}$.
    \STATE Define $\mathcal B_{\mathrm{min}}(u,z)=X(\theta(u),\phi(\hat{z}))$ and set $\Psi=\mathcal B_{\mathrm{min}} \circ\Phi$.
\ENDIF
\STATE Return $\Psi$
\end{algorithmic}
\end{algorithm}

\paragraph{Admissible-radius asymmetry}
We remark that the two admissibility conditions reveal an intrinsic asymmetry between the two bending modes. For a fixed tubular height $\Delta z$, the major-mode bending satisfies $R < R_{\mathrm{maj}}^{+}$, whereas the minor-mode bending satisfies $R > R_{\min}^{-}$. Thus, $\Psi_{\mathrm{maj}}$ has only a bounded admissible radius range, while $\Psi_{\mathrm{min}}$ admits arbitrarily large radii. Consequently, $\Psi_{\mathrm{min}}$ converges to the straight tube as $R\to\infty$. Indeed,
\[
\phi_{\min}(z)=\frac{z}{\sqrt{R^2-1}}=O(R^{-1}),\qquad
\lambda_{\min}(u)=\frac{R+\cos\theta_{\min}(u)}{\sqrt{R^2-1}}=1+O(R^{-1}),
\]
and $\theta_{\min}(u)\to u$, so that $\mathcal{B}_{\min}(u,z)-(R,0,0)\to(\cos u,\,z,\,\sin u)$, which is the original tube $T$ up to a rigid motion. No analogous limit is available for $\Psi_{\mathrm{maj}}$ at fixed $\Delta z$, since its radius is capped by $R^{+}_{\mathrm{maj}}$. This radius-admissibility asymmetry is important in the discrete setting: although both maps are conformal in the smooth sense, the mode with access to arbitrarily gentle bending is expected to introduce smaller chordal discretization error after vertex relocation.

This asymmetry clarifies the practical role of the two bending modes. The minor-mode bending is the preferred choice when a toroidal target is desired with minimal additional discrete distortion, whereas the major-mode bending is useful when a full major-circle wrapping is required by the downstream representation. 

Altogether, the major and minor conformal bending maps provide additional flexibility for reducing overall geometric distortion in the parameterization of tube-like surfaces while preserving conformality. The optimal bending parameters can be determined by minimizing other prescribed distortion measures, such as the area distortion. A detailed analysis is provided in Section~\ref{sect:geometry}.

\begin{figure}[t]
    \centering
    \includegraphics[width=\linewidth]{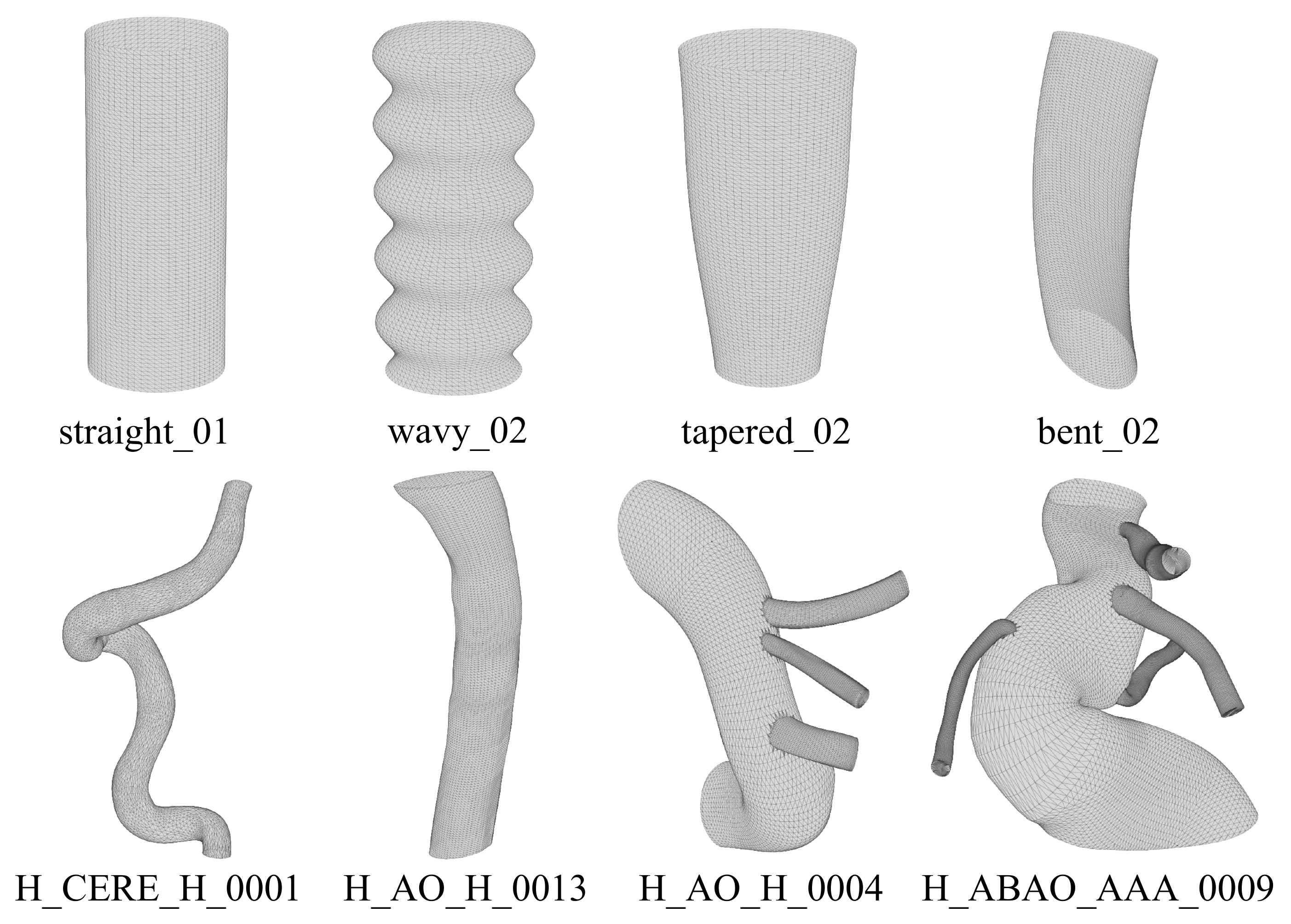}
    \caption{\textbf{Representative meshes in the experimental datasets.} Top: Synthetic tube meshes from the straight, wavy, tapered, and bent geometry families. Bottom: Representative real vascular meshes from the Vascular Model Repository, including both simple single-branch and complex multi-branch geometries.}
    \label{fig:data_illu}
\end{figure}

\begin{figure}[t!]
    \centering
    \includegraphics[width=\linewidth]{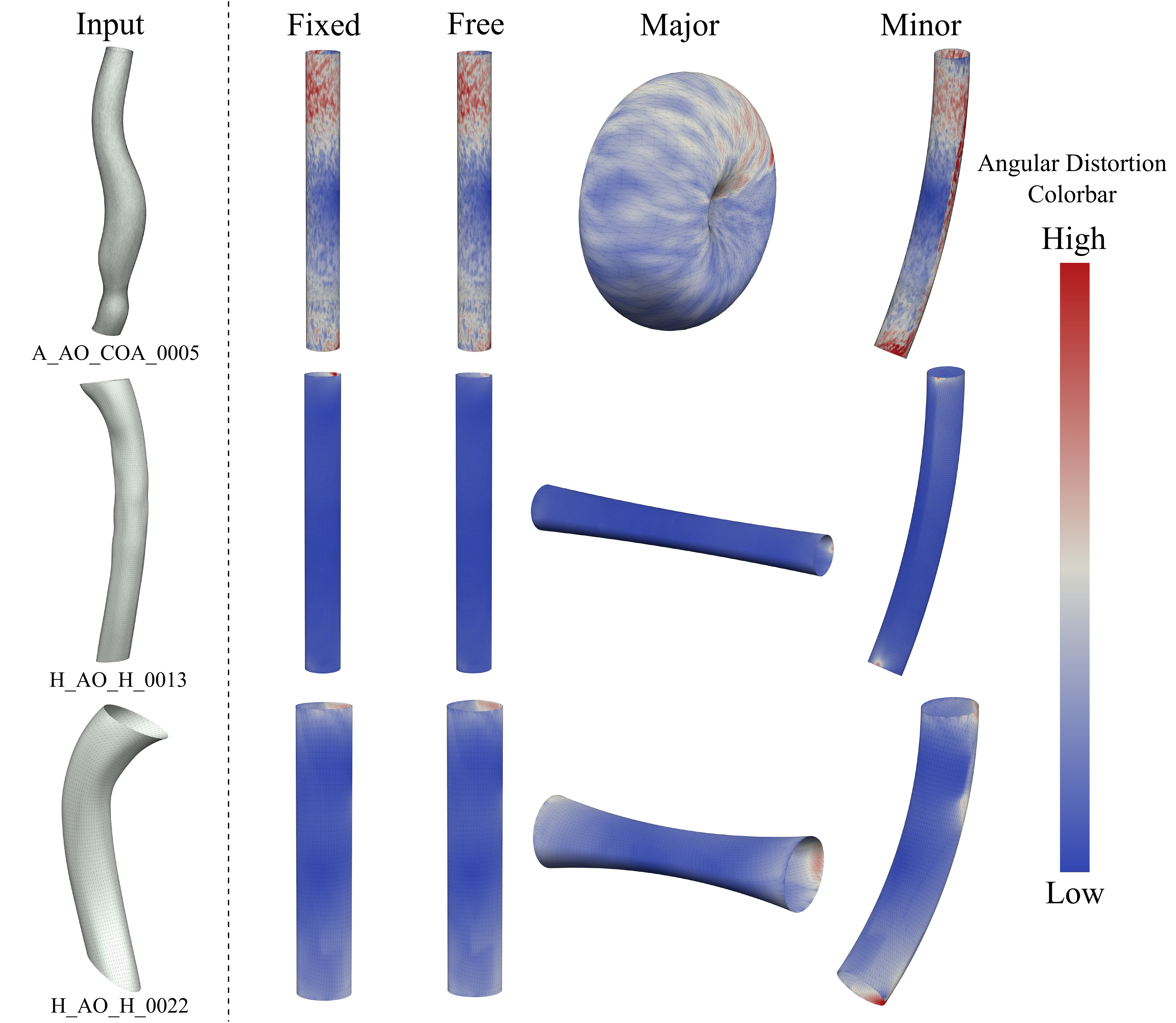}
    \caption{\textbf{Visualization of conformal tubular parameterization and toroidal bending results on vascular surfaces.} From left to right: the input mesh, fixed-boundary conformal tubular parameterization, free-boundary conformal tubular parameterization, major-mode toroidal bending, and minor-mode toroidal bending. Surface colors indicate angular distortion, with blue denoting low distortion and red denoting high distortion. The appearance of $\Psi_{\mathrm{maj}}$ varies across meshes because the tube height and the area-optimal radius $R$ determine its meridional extent: a small extent yields a narrow toroidal band (e.g., H\_AO\_H\_0013), while a large extent covers almost the entire torus (e.g., A\_AO\_COA\_0005).}
    \label{fig:real_result}
\end{figure}

\section{Experiments}\label{sect:experiments}
We evaluate our proposed conformal parameterization methods on both synthetic and real surface meshes with tube topology (see Fig.~\ref{fig:data_illu} and \ref{fig:real_result}). The synthetic benchmark contains $42$ triangulated tube meshes generated procedurally from $4$ geometry families: \texttt{straight}, \texttt{bent}, \texttt{tapered}, and \texttt{wavy}. For each family, we add boundary noise to selected test cases. The real-world dataset consists of 96 tubular vascular structures derived from the Vascular Model Repository~\cite{Wilson2013VascularModelRepository}, including $74$ simple single-branch cases for standard testing and $22$ complex multi-branch cases for stress testing. For each multi-branch case, the two openings of the main vessel are retained as the boundary loops of the parameterized surface, while all remaining branch openings are capped during preprocessing. Hence, all meshes supplied to our method are connected genus-zero surfaces with exactly two boundary components and satisfy the tube-topology assumption. Unless otherwise stated, the combination coefficient in the boundary extension is set to $\tau=0.15$ in all experiments. For the optimization of $(L,s)$, we first minimize over $L\in[0,10]$ with $s=0$ using Brent's method to obtain $L_0$, and then jointly optimize $(L,s)$ over $[0,100]\times[-2\pi,2\pi]$ using Powell's method from $(L_0,0)$, with tolerance $10^{-4}$ and at most $120$ iterations. If the result is non-finite or the objective does not decrease, $(L_0,0)$ is retained. All experiments are implemented in Python using \texttt{NumPy}, \texttt{SciPy}, \texttt{trimesh}, and \texttt{networkx}.

\subsection{Parameter Sensitivity and Ablation Study}
We first evaluate the sensitivity of the proposed framework to three algorithmic parameters: the seam-correction strip width $d$, the cycle-Laplacian smoothing weight $\omega$, and the number of boundary-extension layers $K$. We also perform ablation experiments on the pipeline stages. The reported metric is the mean angular distortion in degrees:
\[
d_{\text{angle}} = \text{mean}\left(\left|\angle [f(v_i), f(v_j), f(v_k)] -  \angle [v_i, v_j, v_k]\right|\right),
\]
where $[v_i, v_j, v_k]$ is a triangle in the original mesh and $f$ is the parameterization. Lower values of $d_{\text{angle}}$ indicate better conformality. As discussed in Section~\ref{subsect:quasi_conformal}, this discrete angle-based measure provides a geometric counterpart to the Beltrami-coefficient-based characterization of local conformal distortion.

\paragraph{Effect of the seam-correction strip width $d$}
Table~\ref{tab:seam_width_counts} indicates that a moderately localized strip gives the most stable behavior. On the synthetic group, the median values are very close across the smaller strip widths, and the narrowest strip attains the best median and the largest number of case-wise wins. This suggests that, for many procedurally generated tubes, the seam-induced error is confined to a very thin neighborhood of the artificial cut. The real vascular meshes are less regular: their local radius variation, nonuniform sampling, and boundary geometry make the seam-affected region less predictable. In this regime, $d=0.20$ gives the best mean distortion, whereas $d=0.25$ gives the lowest median distortion and the largest number of case-wise best results.

This mean--median discrepancy is informative. A wider strip can benefit many individual vascular surfaces by covering a larger seam-affected neighborhood, but it is also more sensitive to high-distortion outliers. Once the strip becomes too wide, the correction may interact with regions that are not primarily seam-dominated, and the fixed boundary condition on the correction strip can force unnecessary redistribution of angular distortion away from the cut. We therefore use $d=0.20$ as the default value in subsequent experiments, since it provides the most robust compromise between covering the seam-affected region and avoiding unnecessary global reparameterization.

\begin{table}[t]
\centering
\scriptsize
\begin{tabular}{|c|ccc|ccc|}
\hline
\multirow{2}{*}{$d$}
& \multicolumn{3}{c|}{Synthetic ($n=42$)}
& \multicolumn{3}{c|}{Real ($n=96$)} \\
\cline{2-7}
& Mean & Median & Best
& Mean & Median & Best \\
\hline
0.05
& 0.3858 & \textbf{0.3488} & \textbf{13}
& 3.5282 & 2.7911 & 1 \\
0.10
& 0.3788 & 0.3491 & 1
& 3.2662 & 2.5927 & 4 \\
0.15
& 0.3742 & 0.3500 & 9
& 3.0745 & 2.3785 & 11 \\
0.20
& \textbf{0.3729} & 0.3504 & 11
& \textbf{2.9573} & 2.0245 & 19 \\
0.25
& 0.3795 & 0.3503 & 7
& 3.4323 & \textbf{1.9176} & \textbf{40} \\
0.30
& 0.4071 & 0.3580 & 1
& 4.7467 & 2.8967 & 21 \\
\hline
\end{tabular}
\caption{Effect of the seam-correction strip width $d$ on the fixed-boundary parameterization. ``Mean'' and ``Median'' denote the whole-surface mean angular distortion in degrees. ``Best'' denotes the number of cases for which a given strip width attains the lowest distortion among all tested widths.}
\label{tab:seam_width_counts}
\end{table}

\paragraph{Effect of the cycle-Laplacian smoothing weight $\omega$}
Table~\ref{tab:smoothing_weight} separates the synthetically boundary-noisy cases from the remaining meshes. In the boundary-clean group, the raw extension already gives the best aggregate statistics, and the differences among smoothing weights are small. This indicates that smoothing is not the dominant factor when the original boundary loops are sufficiently compatible with the auxiliary extension.

The boundary-noisy subset shows a different behavior. Moderate smoothing gives the best average performance, whereas very strong smoothing gives the best median and the largest number of case-wise wins. This indicates that strong smoothing can suppress high-frequency boundary oscillations in many typical cases, but may also over-regularize the artificial ring and create outliers when the extension geometry becomes poorly aligned with the local surface. The smoothing term therefore acts as a regularizer for the extension scaffold, not as a direct conformal optimization of the original surface.

Overall, the experiment suggests that cycle-Laplacian smoothing should be used as a stabilizing regularizer rather than as an aggressive optimization step. For inputs without artificial boundary perturbation, little or no smoothing is already sufficient. For noisy boundaries, smoothing is useful, and $\omega=0.50$ provides the best compromise between average distortion reduction and robustness against outliers. We therefore use $\omega=0.50$ as the default smoothing weight in the subsequent experiments.

\begin{table}[t]
\centering
\scriptsize
\begin{tabular}{|c|ccc|ccc|}
\hline
\multirow{2}{*}{Setting}
& \multicolumn{3}{c|}{Boundary Clean ($n=108$)}
& \multicolumn{3}{c|}{Boundary Noisy ($n=30$)} \\
\cline{2-7}
& Mean & Median & Best
& Mean & Median & Best \\
\hline
Raw $(\omega=0)$
& \textbf{2.6673} & \textbf{1.8089} & \textbf{48}
& 0.3689 & 0.3355 & 2 \\
$\omega=0.25$
& 2.6768 & 1.8092 & 12
& 0.3686 & 0.3327 & 1 \\
$\omega=0.50$
& 2.6753 & 1.8094 & 11
& \textbf{0.3639} & 0.3319 & 3 \\
$\omega=0.75$
& 2.6770 & 1.8106 & 5
& 0.3841 & 0.3315 & 3 \\
$\omega=1.00$
& 2.6781 & 1.8105 & 3
& 0.3839 & 0.3313 & 1 \\
$\omega=1.25$
& 2.6716 & 1.8105 & 1
& 0.3890 & 0.3272 & 2 \\
$\omega=1.50$
& 2.6696 & 1.8104 & 28
& 0.3889 & \textbf{0.3270} & \textbf{18} \\
\hline
\end{tabular}
\caption{Effect of the cycle-Laplacian smoothing weight $\omega$ in the free-boundary parameterization with one extension layer. ``Raw'' denotes the unsmoothed boundary extension, corresponding to $\omega=0$. ``Mean'' and ``Median'' denote the whole-surface mean angular distortion in degrees. ``Best'' denotes the number of cases for which a given smoothing weight attains the lowest distortion among all tested settings. The boundary-clean subset consists of the clean synthetic cases and all real vascular cases, while the boundary-noisy subset consists of the noisy synthetic cases.}
\label{tab:smoothing_weight}
\end{table}

\paragraph{Effect of the number of boundary-extension layers $K$}
Table~\ref{tab:extension_layers} shows that one extension layer is sufficient for the boundary-clean group. In this regime, the main benefit of boundary extension is already achieved once the prescribed circular constraints are moved away from the original boundary loops. Adding more layers gives the auxiliary extension more influence over the optimized tube height, the longitudinal coordinate distribution, and the subsequent correction steps, without providing a consistent improvement.

For noisy boundaries, deeper extension can improve the median behavior, which is consistent with the idea that moving the circular constraints farther away reduces the direct influence of unreliable boundary samples. However, the mean distortion and case-wise best counts show that this advantage is not uniformly robust. Additional layers can help when the extension forms a smooth continuation of the surface, but may become harmful when the generated extension is poorly conditioned.

Therefore, the parameter $K$ controls a tradeoff between boundary relaxation and extension stability. A deeper extension can make the original boundary freer, but it also increases the geometric and numerical influence of artificial triangles that are not part of the target surface. We use $K=1$ in the subsequent experiments as the most stable default setting.

\begin{table}[t]
\centering
\scriptsize
\begin{tabular}{|c|ccc|ccc|}
\hline
\multirow{2}{*}{$K$}
& \multicolumn{3}{c|}{Boundary Clean ($n=108$)}
& \multicolumn{3}{c|}{Boundary Noisy ($n=30$)} \\
\cline{2-7}
& Mean & Median & Best
& Mean & Median & Best \\
\hline
1
& \textbf{2.6753} & \textbf{1.8094} & \textbf{60}
& \textbf{0.3639} & 0.3319 & \textbf{11} \\
2
& 2.6822 & 1.8278 & 6
& 0.3929 & 0.3164 & \textbf{11} \\
3
& 2.7366 & 1.8343 & 42
& 0.3815 & \textbf{0.3036} & 8 \\
\hline
\end{tabular}
\caption{Effect of the number of boundary-extension layers $K$ under different boundary-quality groups. 
The smoothing weight is fixed at $\omega=0.5$. 
``Mean'' and ``Median'' denote the whole-surface mean angular distortion in degrees. 
``Best'' denotes the number of cases for which a given layer number attains the lowest distortion among all tested settings. }
\label{tab:extension_layers}
\end{table}

\paragraph{Pipeline ablation} Table~\ref{tab:pipeline_ablation} reports a sequential-state ablation rather than a standard leave-one-module-out ablation. This choice is intentional because the stages in our pipeline are not independent modules. For example, removing the seam correction process will result in the same parallelogram in both the initial tube and the following interior refinement, since the boundary shape remains unchanged. Therefore, we evaluate the distortion after each meaningful intermediate state: the initial fixed-boundary map, the seam-corrected map, the interior-refined fixed-boundary map, and the free-boundary map. This allows us to identify whether each stage acts as a standalone distortion reducer or mainly prepares a better configuration for the next optimization step.

On clean synthetic meshes, the initial map is already highly competitive and attains most case-wise wins, indicating that little correction is needed when the geometry and boundary constraints are well aligned. On noisy synthetic meshes, the free-boundary pipeline is clearly dominant, showing that boundary extension effectively reduces boundary-induced bias rather than merely smoothing the result. The seam-only result is not consistently better than the initial map, which supports the interpretation that the localized strip solve mainly repairs the cut neighborhood and may introduce transition distortion before the interior degrees of freedom are re-optimized. On real single-branch vascular meshes, the interior-refined fixed-boundary pipeline performs best, suggesting that the dominant residual error is interior conformality rather than boundary bias. For real multi-branch cases, the free-boundary pipeline becomes preferable, reflecting the stronger influence of irregular trimming loops, nonuniform sampling, and branch-induced geometric complexity. Overall, the ablation indicates that the fixed-boundary pipeline is sufficient for boundary-compatible inputs, while the free-boundary formulation is most useful when boundary irregularity or multi-branch complexity becomes the main source of distortion.

\begin{table}[t]
\centering
\scriptsize
\resizebox{\textwidth}{!}{%
\begin{tabular}{|l|ccc|ccc|ccc|ccc|}
\hline
\multirow{3}{*}{Pipeline}
& \multicolumn{6}{c|}{Synthetic}
& \multicolumn{6}{c|}{Real} \\
\cline{2-13}
& \multicolumn{3}{c|}{Clean ($n=12$)}
& \multicolumn{3}{c|}{Noisy ($n=30$)}
& \multicolumn{3}{c|}{Single ($n=74$)}
& \multicolumn{3}{c|}{Multiple ($n=22$)} \\
\cline{2-13}
& Mean & Median & Best
& Mean & Median & Best
& Mean & Median & Best
& Mean & Median & Best \\
\hline
Initial
& 0.2899 & 0.2554 & \textbf{9}
& 0.4376 & 0.3723 & 0
& 3.7559 & 3.3411 & 1
& 4.0201 & 2.4295 & 0 \\
Seam
& 0.3075 & 0.2703 & 0
& 0.4426 & 0.3716 & 0
& 4.2723 & 3.7910 & 0
& 4.3308 & 2.3954 & 1 \\
Interior
& 0.2952 & 0.2602 & 0
& 0.4039 & 0.3611 & 1
& \textbf{2.8439} & \textbf{2.0245} & \textbf{46}
& 3.3388 & 2.0138 & 5 \\
Free
& \textbf{0.2842} & \textbf{0.2458} & 3
& \textbf{0.3639} & \textbf{0.3319} & \textbf{29}
& 2.8715 & 2.1269 & 27
& \textbf{3.3193} & \textbf{2.0048} & \textbf{16} \\
\hline
\end{tabular}
}
\caption{Ablation study of the proposed tubular parameterization pipeline, grouped by synthetic and real subsets. ``Mean'' and ``Median'' denote the whole-surface mean angular distortion in degrees. ``Best'' denotes the number of cases for which a given setting attains the lowest distortion among all tested settings.}
\label{tab:pipeline_ablation}
\end{table}

\subsection{Comparison between Conformal Solvers for Tubular Parameterization}
Note that in our proposed computational pipelines for fixed-boundary and free-boundary conformal tubular parameterization, we primarily utilize the LBS solver~\cite{Lui2013beltrami} for computing a conformal map on the plane and performing the seam correction. It is natural to ask whether the LBS solver can be replaced with other existing conformal mapping solvers in the proposed tubular parameterization framework. Here, we perform an extensive comparison between the LBS-based methods and two existing conformal solvers, namely holomorphic differentials (HD)~\cite{gu2002computing} and discrete Ricci flow (RF)~\cite{zeng2013ricci}. Since both HD and RF produce planar parameterizations, we first identify a cut seam for each input mesh using the same detection procedure and then apply the respective conformal mapping solvers. Finally, we apply the same conformal lifting procedure to all planar mapping results to obtain the corresponding tubular parameterizations for a fair comparison. The benchmark results on synthetic and real meshes are reported in Tables~\ref{tab:baseline_synthetic} and~\ref{tab:baseline_real}.

\begin{table}[t]
\centering
\scriptsize
\setlength{\tabcolsep}{3.5pt}
\resizebox{\textwidth}{!}{$
\begin{tabular}{|c|ccc|ccc|c|c|}
\hline
\multirow{2}{*}{Method}
& \multicolumn{3}{c|}{Boundary Clean ($n=12$)}
& \multicolumn{3}{c|}{Boundary Noisy ($n=30$)}
& \multirow{2}{*}{Flip Ratio}
& \multirow{2}{*}{Time (s)} \\
\cline{2-7}
& Mean & Median & Best
& Mean & Median & Best
& & \\
\hline
LBS-based mapping (Fixed boundary)
& 0.2952 & 0.2602 & 0
& 0.4039 & 0.3611 & 0
& \textbf{0.000000} \% & 0.5759 \\

LBS-based mapping (Free boundary)
& \textbf{0.2842} & \textbf{0.2458} & 3
& \textbf{0.3639} & \textbf{0.3319} & \textbf{23}
& \textbf{0.000000} \% & 0.6454 \\

HD-based mapping
& 0.2881 & 0.2556 & 4
& 0.3688 & 0.3510 & 0
& 0.000392 \% & \textbf{0.2632} \\

RF-based mapping
& 0.2871 & 0.2550 & \textbf{8}
& 0.3641 & 0.3441 & 7
& \textbf{0.000000} \% & 1.4252 \\
\hline
\end{tabular}
$}
\caption{Comparison between using different conformal solvers in our proposed tubular parameterization framework on synthetic meshes. Here, ``Mean'' and ``Median'' denote the mean and median, respectively, of the per-mesh whole-surface mean angular distortion of the overall tubular parameterization in degrees. ``Best'' denotes the number of cases for which a method attains the lowest distortion among all four methods; ties are credited to all tied methods. ``Flip Ratio'' denotes the average ratio of flipped faces over all 42 synthetic meshes, and ``Time'' denotes the average runtime over the synthetic benchmark.}
\label{tab:baseline_synthetic}
\end{table}

\begin{table}[t!]
\centering
\scriptsize
\setlength{\tabcolsep}{3.5pt}
\resizebox{\textwidth}{!}{$
\begin{tabular}{|c|ccc|ccc|c|c|}
\hline
\multirow{2}{*}{Method}
& \multicolumn{3}{c|}{Single ($n=74$)}
& \multicolumn{3}{c|}{Multiple ($n=22$)}
& \multirow{2}{*}{Flip Ratio}
& \multirow{2}{*}{Time (s)} \\
\cline{2-7}
& Mean & Median & Best
& Mean & Median & Best
& & \\
\hline
LBS-based mapping (Fixed boundary)
& 2.8439 & 2.0245 & 0
& 3.3388 & 2.0138 & 1
& 0.001752 \% & 2.4558 \\

LBS-based mapping (Free boundary)
& 2.8715 & 2.1269 & 0
& \textbf{3.3193} & 2.0048 & \textbf{10}
& \textbf{0.001229} \% & 2.6788 \\

HD-based mapping
& 0.3248 & 0.2604 & 0
& 4.7392 & \textbf{1.9003} & 9
& 0.062876 \% & \textbf{1.1137} \\

RF-based mapping
& \textbf{0.2963} & \textbf{0.2193} & \textbf{74}
& 19.6012 & 15.3562 & 2
& 3.829500 \% & 5.0956 \\
\hline
\end{tabular}
$}
\caption{Comparison between using different conformal solvers in our proposed tubular parameterization framework on real vascular surfaces. Here, ``Mean'' and ``Median'' denote the mean and median, respectively, of the per-mesh whole-surface mean angular distortion of the overall tubular parameterization in degrees. ``Best'' denotes the number of cases for which a method attains the lowest distortion among all four methods. ``Flip Ratio'' denotes the average ratio of flipped faces over all 96 real meshes, and ``Time'' denotes the average runtime over the real benchmark.}
\label{tab:baseline_real}
\end{table}

\begin{figure}[t]
    \centering
    \includegraphics[width=\linewidth]{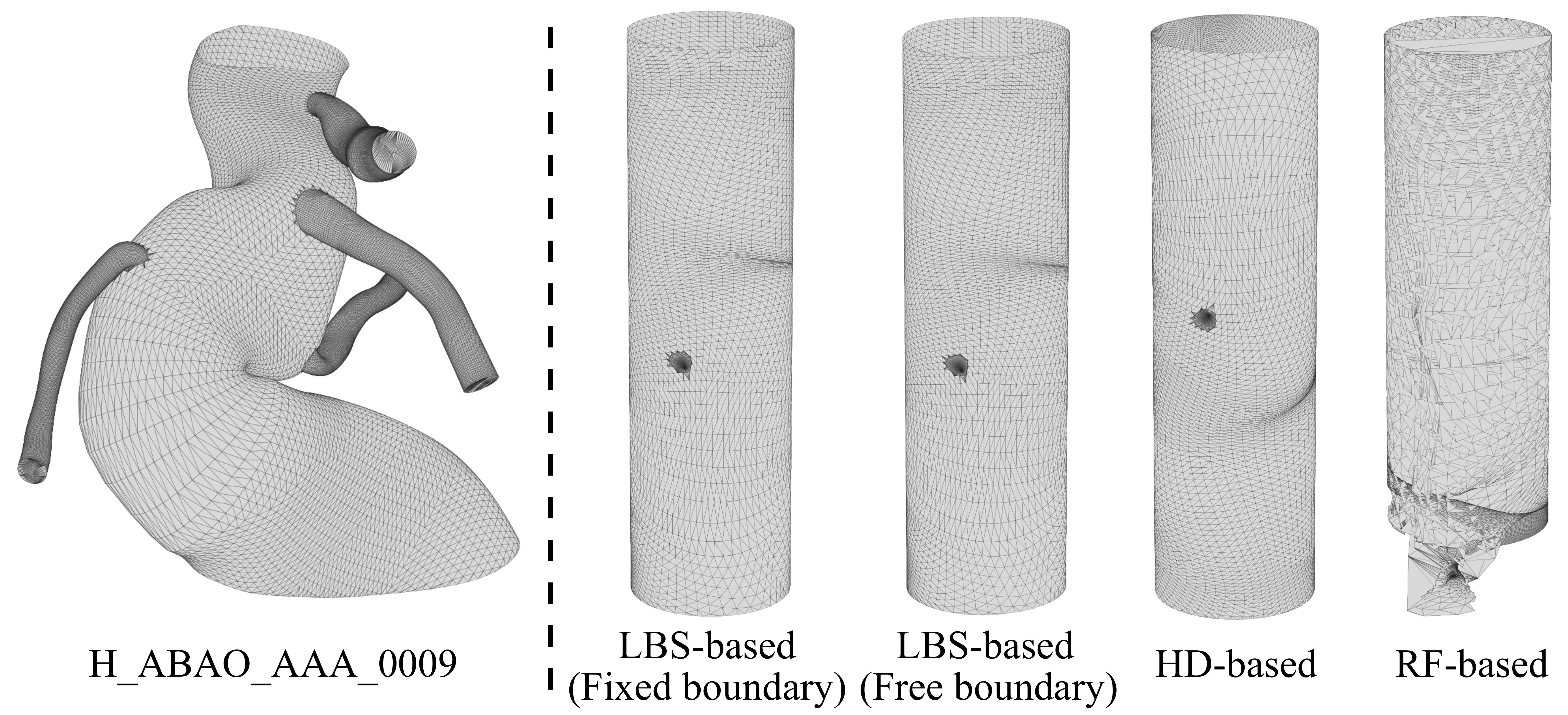}
    \caption{\textbf{Conformal tubular parameterizations of a challenging multi-branch vascular surface.} From left to right: the input mesh and the tubular parameterizations produced by our proposed framework together with the LBS-based fixed-boundary and free-boundary methods, holomorphic differentials (HD), and discrete Ricci flow (RF). It can be observed that the LBS-based methods and HD preserve a coherent tubular parameterization, whereas RF exhibits severe geometric degeneration and local collapse.}
    \label{fig:baseline_comparison}
\end{figure}

On the synthetic meshes, all four methods (LBS with fixed boundary, LBS with free boundary, HD, and RF) operate in a broadly similar accuracy regime. The free-boundary LBS-based method achieves the best aggregate mean and median distortions, as well as the largest total number of case-wise wins, while HD and RF remain competitive and attain the best result on a non-negligible subset of cases. The fixed-boundary LBS-based method yields slightly higher distortion than the other methods. This modest degradation is consistent with its direct imposition of circular constraints on the original boundary loops, which reduces the available boundary degrees of freedom and allows a mismatch between the input boundaries and the prescribed tubular target to influence the interior parameterization. The small gap on the clean synthetic meshes indicates that such constraints are not intrinsically detrimental when the boundaries are regular and compatible with the target domain. In contrast, the larger gap on the boundary-noisy subset shows that the fixed-boundary formulation is more sensitive to boundary perturbations, whereas the extension used in the free-boundary method provides an auxiliary region for absorbing the boundary--target mismatch. Overall, the differences among the solvers remain relatively small across the synthetic benchmark, indicating that none of these methods uniformly dominates every instance. Their relative performance is instead determined mainly by how each formulation handles the more challenging boundary configurations.

Evaluation on the real vascular meshes reveals a stronger separation between simple and geometrically complex inputs. On all 74 simple single-branch surfaces, RF produces the lowest angular distortion, and HD is also markedly more conformal than either of the LBS-based methods. A different picture emerges on the 22 complex multi-branch cases. The free-boundary method gives the lowest mean error and the largest number of case-wise wins, whereas HD gives the lowest median and nearly the same number of wins. The low median of HD shows that it remains highly effective on a typical complex example. However, the large separation between its mean and median indicates that it produces substantially larger errors on a small set of difficult cases. By contrast, the closer mean--median behavior of the LBS-based methods indicates a more controlled upper tail. In other words, HD provides stronger typical-case conformality, while the proposed tubular formulation sacrifices some best-case accuracy in exchange for greater robustness on the hardest geometries. RF, despite its best performance on the single-branch surfaces, deteriorates markedly on the multi-branch subset, with both its mean and median distortion several times larger than those of the other three methods. Fig.~\ref{fig:baseline_comparison} shows a representative example: the LBS-based methods and HD preserve a coherent tubular parameterization, whereas RF exhibits severe geometric degeneration and local collapse.

The flip ratio reveals a further distinction among the four methods. For each triangle $f$, we compute its signed area 
\[a_f=\frac{1}{2}\det(q_1-q_0,\,q_2-q_0)\]
in the tubular parameter coordinates $q_i=(u_i,z_i)$, where $u_1,u_2$ are shifted by integer multiples of $2\pi$ into $[u_0-\pi,\,u_0+\pi)$ to account for the periodicity. After excluding numerically degenerate triangles with $|a_f|\le 10^{-12}$, we take the majority sign as the global orientation, and the flip ratio as the percentage of triangles with the opposite sign. Both the fixed- and free-boundary LBS-based methods exhibit negligible flip ratios, on the order of $0.001\%$, on real meshes. In the continuous setting, a quasi-conformal map with $\|\mu\|_\infty<1$ is a homeomorphism under appropriate normalization~\cite{Ahlfors1960mapping}. In the discrete setting, however, bijectivity of the LBS solution is not guaranteed in general. A sufficient condition is that every interior vertex is a strictly convex combination of its neighbors, i.e., all edge weights of the discrete operator are positive, and that the boundary is mapped homeomorphically onto the boundary of a convex domain; the resulting piecewise linear map is then one-to-one~\cite{Tutte1963DrawGraph,floater2003one}. These conditions are not always satisfied in our pipeline: the anisotropic cotangent-type weights may become negative on obtuse or highly stretched triangles, the correction strip $\mathcal{A}_d$ is generally non-convex, and the parallelogram step involves periodic rather than purely Dirichlet constraints. Consequently, a small number of flipped faces may occur, particularly on extremely slender triangles. HD is less stable in this respect, although its flip ratio remains relatively limited. RF, by contrast, yields a flip ratio more than 1000 times that of the LBS-based mappings. Its deterioration in complex cases therefore reflects not only a loss of conformality but also a substantial loss of mapping validity. Consequently, a low distortion value in the easier cases does not, by itself, establish robustness: for geometrically challenging inputs, the ability to suppress orientation reversals becomes an equally important criterion.

In terms of computational cost, HD is the fastest and RF is the most expensive, while the LBS-based methods lie between them. The boundary-extension procedure adds only a modest overhead relative to the fixed-boundary pipeline.

Altogether, the choice of the conformal solver to be used in our proposed tubular parameterization framework is primarily determined by the geometric regime rather than by runtime: the HD and RF methods already work well on simple surfaces, while the fixed-boundary LBS-based method provides explicit tubular coordinates for boundary-compatible inputs and the free-boundary LBS-based method offers the most reliable compromise for noisy boundaries and geometrically complex multi-branch surfaces. For a balance among conformality, efficiency, and applicability to complex structures, the LBS-based solver is adopted as the default choice in our proposed tubular parameterization framework. The user may also replace it with other conformal solvers for different specific applications while keeping the other components in the pipeline unchanged.

\subsection{Geometric Compatibility of Target Domains and the Role of Toroidal Bending} \label{sect:geometry}

\begin{figure}[t]
    \centering
    \includegraphics[width=\linewidth]{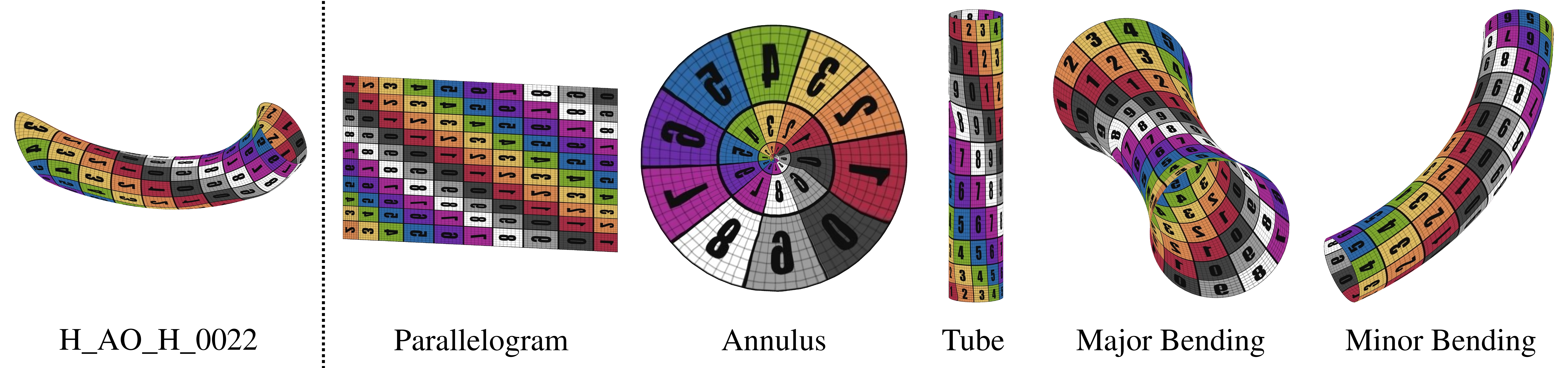}
    \caption{\textbf{Visual comparison of different parameterization target domains.} From left to right: the input vascular surface and its parameterizations onto a parallelogram, an annulus, a straight tube, and the major- and minor-mode toroidal targets. The same texture is transferred to all target domains to illustrate their geometric compatibility and the resulting deformation patterns.}
    \label{fig:geo_compa}
\end{figure}

We next evaluate the influence of the proposed toroidal bending maps. Since the bending construction can be applied to any tubular shape, the specific method used to obtain the tubular parameterization is not the primary concern in this experiment. We therefore isolate the effect of the final target-domain geometry by applying the same parameterization pipeline in all cases. Specifically, we apply the LBS-based solver to obtain a conformal parameterization and compare five target domains (see Fig.~\ref{fig:geo_compa}), namely a parallelogram, an annulus, a straight tube, and the major- and minor-mode toroidal bendings, in terms of the resulting angular and area distortions. Here, the toroidal bending radius $R$ is determined automatically by minimizing the area distortion given by
\[
  d_{\text{area}}=
  \sqrt{\frac{1}{|F|}\sum_{i=1}^{|F|}\left(\log\left(
  \frac{A_i^\prime/\sum_j A_j^\prime}
  {A_i/\sum_j A_j}
  \right)\right)^2},
\]
where $F$ is the set of all triangles of the input surface and $A_i,A_i^\prime$ are the areas of the $i$-th corresponding triangles of the input surface and the final parameterization, with $i = 1, 2, \dots, |F|$. Note that the factors $\sum_j A_j$ and $\sum_j A_j^\prime$ serve as normalization factors so that $d_{\text{area}} = 0$ if and only if all triangle areas are preserved (up to a uniform rescaling) in the parameterization. Table~\ref{tab:target_domain} reports the comparison results.

\begin{table}[t]
\centering
\scriptsize
\setlength{\tabcolsep}{3.5pt}
\begin{tabular}{|c|ccc|ccc|ccc|}
\hline
\multirow{2}{*}{Target}
& \multicolumn{3}{c|}{Synthetic ($n=42$)}
& \multicolumn{3}{c|}{Single ($n=74$)}
& \multicolumn{3}{c|}{Multiple ($n=22$)} \\
\cline{2-10}
& Mean & Median & Best
& Mean & Median & Best
& Mean & Median & Best \\
\hline

\multicolumn{10}{|c|}{Angular Distortion} \\
\hline

Parallelogram
& 0.3938 & \textbf{0.3468} & 14
& 3.7575 & 3.3450 & 1
& 4.0204 & 2.4284 & 2 \\

Annulus
& 1.9910 & 1.9318 & 0
& 3.5083 & 2.8638 & 0
& 3.5761 & 2.2309 & 0 \\

Tube
& \textbf{0.3728} & 0.3504 & \textbf{24}
& \textbf{2.8439} & \textbf{2.0245} & \textbf{37}
& \textbf{3.3388} & \textbf{2.0138} & \textbf{14} \\

$\Psi_{\mathrm{maj}}$
& 1.1817 & 1.1606 & 1
& 2.9184 & 2.0699 & 18
& 3.4386 & 2.0730 & 1 \\

$\Psi_{\mathrm{min}}$
& 0.3945 & 0.3653 & 3
& 2.8449 & 2.0296 & 18
& 3.3487 & 2.0489 & 5 \\

\hline
\multicolumn{10}{|c|}{Area Distortion} \\
\hline

Parallelogram
& 0.2570 & \textbf{0.2552} & 9
& 0.4567 & 0.4073 & 4
& 10.5649 & 9.9064 & 0 \\

Annulus
& 5.0514 & 4.7586 & 0
& 15.3271 & 13.5998 & 0
& 19.7253 & 13.9929 & 0 \\

Tube
& \textbf{0.2567} & 0.2555 & \textbf{21}
& 0.4392 & 0.3874 & 5
& 10.5659 & 9.9010 & 0 \\

$\Psi_{\mathrm{maj}}$
& 1.1887 & 1.2152 & 2
& 1.2633 & 0.5464 & \textbf{40}
& 10.3058 & 10.0111 & 6 \\

$\Psi_{\mathrm{min}}$
& 0.2824 & 0.2710 & 10
& \textbf{0.4146} & \textbf{0.3793} & 25
& \textbf{10.1343} & \textbf{9.5304} & \textbf{16} \\

\hline
\end{tabular}
\caption{Comparison of different target-domain geometries on the synthetic and real meshes. ``Mean'' and ``Median'' denote the mean and median, respectively, of the per-mesh whole-surface mean distortion. Angular distortion is measured in degrees. ``Best'' denotes the number of cases for which a target attains the lowest distortion among all five targets.}
\label{tab:target_domain}
\end{table}

From the experimental results, it can be observed that the straight tube provides the most stable angular-distortion baseline. Specifically, it achieves the lowest mean angular distortion and the largest number of case-wise wins on all three datasets. The minor-mode bending remains particularly close to this baseline. Its average angular penalty relative to the tube is only approximately $0.001^\circ$ on the single-branch meshes and $0.010^\circ$ on the multi-branch meshes. This supports the theoretical interpretation that the minor mode can realize a gentle toroidal deformation while largely preserving the conformal structure of the original tube coordinates. The major mode introduces a more visible angular penalty, especially on the synthetic examples, consistent with the fact that its radius is bounded above by $R^{+}_{\mathrm{maj}}$ for a fixed tube height, so that it cannot be made arbitrarily close to the straight tube.

The planar targets further show that topological compatibility alone is insufficient for geometric fitting. The parallelogram is competitive on the synthetic meshes because it differs intrinsically from the straight tube only through cutting and wrapping. However, it retains an artificial seam and does not directly encode the periodic circumferential structure. The annulus produces no case-wise area win in any dataset and is consistently ranked last in area distortion on all real meshes. Although its angular distortion on the real datasets is not as severe as its area distortion, its highly nonuniform planar scaling makes it unsuitable as a direct geometric representation of an elongated tubular surface. Its main utility in the proposed framework is therefore as an auxiliary computational domain rather than as the final embedding.

The two toroidal bending modes display a marked asymmetry. The angular distortion of $\Psi_{\mathrm{min}}$ closely follows that of the straight tube. On the multi-branch dataset, the absolute case-wise difference between the two targets remains very small across all meshes. This confirms that the minor-mode bending preserves the conformal structure of the tubular coordinates to high accuracy, and that the remaining discrepancy is primarily a discrete chordal effect rather than a change in the underlying smooth conformal map. By contrast, $\Psi_{\mathrm{maj}}$ introduces a more visible angular penalty on the synthetic meshes and a heavier upper tail on the real meshes. This difference agrees with the admissible-radius analysis in Section~\ref{sect:conformal_bending}: $\Psi_{\mathrm{min}}$ converges to the straight tube as $R$ increases, whereas the radius of $\Psi_{\mathrm{maj}}$ is upper-bounded for a fixed tube height and the bending therefore cannot be made arbitrarily gentle.

The area distortion results further reveal that the usefulness of toroidal bending is geometry-dependent. On the synthetic meshes, the straight tube is generally preferable because the procedural surfaces were generated around an approximately axial tubular model. In this regime, toroidal bending introduces geometric variation that is not required by the input. The real single-branch meshes exhibit a less uniform pattern. The major-mode target achieves the most case-wise area-distortion wins, indicating that a full major-circle wrapping can fit certain globally curved surfaces well. However, the discrepancy between its case-wise win count and its average distortion reveals a small set of severe failures. The major mode should therefore be regarded as a specialized target whose effectiveness depends strongly on whether the input geometry is compatible with the prescribed full wrapping. The minor mode yields fewer isolated wins on this subset but exhibits substantially more stable area-distortion behavior while remaining almost indistinguishable from the tube in angular distortion.

The advantage of the minor-mode target becomes unambiguous on the real multi-branch meshes. Its area distortion is lower than that of the straight tube in all 22 cases, and it yields the best area result among the five target domains in 16 cases. At the same time, its angular distortion remains close to the tube baseline. Consequently, $\Psi_{\mathrm{min}}$ best balances the angular and area distortion for every multi-branch example: the straight tube may have marginally lower angular distortion in some cases, but it does not dominate the minor torus once geometric area fitting is also considered. This systematic behavior indicates that the additional degree of freedom in curvature is particularly useful when a single straight axis is inadequate to represent the global organization of the surface.

These observations clarify the role and necessity of toroidal bending in the proposed parameterization framework for tubular surfaces. While the straight tube remains the natural baseline for nearly cylindrical inputs, the toroidal bending maps can further expand the admissible family of topology-preserving target geometries, allowing the parameter domain to accommodate global curvature and hence further reduce the overall geometric distortion, without discarding its longitudinal and periodic circumferential coordinates. Among the two constructions, $\Psi_{\mathrm{min}}$ is the preferred general-purpose toroidal target because it provides this geometric adaptability with controlled additional discrete distortion. The major-mode target remains useful when full wrapping along the major circle is explicitly required by the downstream representation.

 \vspace{-3mm}

\section{Conclusion and Discussion}\label{sect:conclusion}

In this work, we developed a conformal parameterization framework for tube-like surfaces together with conformal bending constructions onto toroidal geometries. For open surfaces with two boundary loops, the fixed-boundary pipeline consists of an initial tubular parameterization, a localized quasi-conformal correction around the cut seam, and a final interior refinement. To reduce the influence of noisy or irregular input boundaries, we further introduced a free-boundary variant based on boundary extension, cycle-Laplacian smoothing, and restriction to the original surface. Building on the resulting tube coordinates, we derived two conformal bending maps that provide topology-preserving toroidal target geometries. Altogether, the proposed conformal tubular parameterization and toroidal bending methods provide a suite of computational tools for handling tube-like surfaces with different geometries. As demonstrated by their efficiency and conformality, the proposed methods can be effectively applied to different shape processing and analysis tasks for tubular surfaces such as surface registration and remeshing, in which the conformality of the parameterizations largely facilitates the preservation of local geometric details.

It is noteworthy that the dominant source of parameterization distortion depends on the geometric characteristics of the input surface. For boundary-compatible synthetic surfaces, the initial fixed-boundary map is already highly competitive. The localized seam correction primarily repairs distortion near the artificial cut and prepares a better configuration for the subsequent interior refinement, rather than serving as a consistent standalone reducer of whole-surface distortion. Interior refinement provides the main improvement on many real single-branch vascular surfaces, whereas the free-boundary formulation is most beneficial for noisy boundaries and geometrically complex multi-branch inputs. In terms of the conformal solver to be used in our tubular mapping framework, we note that classical methods such as the holomorphic differentials and discrete Ricci flow can already achieve a low angular distortion on simple surfaces, while the LBS-based methods provide more controlled behavior and substantially fewer orientation reversals on challenging multi-branched geometries.

Also, while the straight tube remains the most stable target in terms of angular distortion, our experiments show that toroidal bending, which leaves the conformal structure of the tubular parameterization unchanged, can further reduce area distortion by adapting the extrinsic geometry of the target domain. Among the two bending constructions, the minor-mode map generally stays close to the straight-tube baseline while providing additional geometric flexibility and improved area fitting, particularly on multi-branch vascular surfaces. This behavior is consistent with its convergence to the straight tube as the major radius increases. The major-mode map is more geometry-dependent and is most appropriate when full wrapping along the major circle is explicitly required. Altogether, combining our proposed tubular parameterization framework and toroidal bending maps allows us to handle a wide range of tube-like surfaces. 

The proposed framework assumes surfaces with exactly two boundary loops. For the multi-branch vascular surfaces in our experiments, this assumption is satisfied by capping all side-branch openings, as described in Section~\ref{sect:experiments}. Such capping, however, comes at a cost: each capped branch becomes a protrusion of the tubular surface, which a conformal map necessarily compresses into a small region of the parameter domain (see Fig.~\ref{fig:baseline_comparison}). This explains the substantially larger area distortion observed on the multi-branch meshes compared with the single-branch ones (see Table~\ref{tab:target_domain}), even though the angular distortion remains low. For highly branched structures such as bronchial trees, these difficulties may be further amplified, since the number of branches is large and the choice of the two retained boundary loops is no longer canonical. A more suitable strategy for such geometries would be to decompose the surface at bifurcations into tube segments, parameterize each segment with the proposed method, and enforce consistency across the junctions, or to combine the tubular formulation with conformal parameterization methods for multiply-connected surfaces~\cite{Choi2021multiconnect}. A more systematic treatment of branching geometries is left for future work.

Several directions remain for future work. First, it would be useful to develop an adaptive strategy for selecting between the fixed- and free-boundary formulations, as well as between straight-tube and toroidal target geometries, based on boundary quality and global geometric compatibility. Second, although the current implementation is already practical for moderate-resolution meshes, further acceleration of the core parameterization stage would be valuable. Moreover, the present framework treats tubular surfaces only; extending it to volumetric tubular domains, such as the solid interior of vessels, is another interesting direction. Finally, it would be interesting to extend the framework to more general surface geometries and topologies for different downstream applications in geometry processing, medical shape analysis, and surface remeshing.

\bibliographystyle{ieeetr}
\bibliography{reference.bib}

\end{document}